# Philosophy Enters the Optics Laboratory:
# Bell's Theorem and its First Experimental Tests (1965-1982)[#]

Olival Freire Jr.

## Abstract

This paper deals with the ways that the issue of completing quantum mechanics was brought into laboratories and became a topic in mainstream quantum optics. It focuses on the period between 1965, when the Bell published what we now call Bell's theorem, and 1982, when Aspect published the results of his experiments. Discussing some of those past contexts and practices, I show that factors in addition to theoretical innovations, experiments, and techniques were necessary for the flourishing of this subject, and that the experimental implications of Bell's theorem were neither suddenly recognized nor quickly highly regarded by physicists. Indeed, I will argue that what was considered good physics after Aspect's 1982 experiments was once considered by many a philosophical matter instead of a scientific one, and that the path from philosophy to physics required a change in the physics community's attitude about the status of the foundations of quantum mechanics.

---

**Introduction**

Quantum non-locality, or entanglement, that is the quantum correlations between systems (photons, electrons, etc) that are spatially separated, is the key physical effect in the burgeoning and highly funded search for quantum cryptography and computation. This effect emerged in relation to the investigation of the possibility of completing quantum theory with supplementary variables, an issue once considered very marginal in physics research. This paper deals with the ways that the issue of completing quantum mechanics, especially completing it according to the criterion of locality, was brought into laboratories and, later on, became a topic in mainstream quantum optics. Discussing some of the past contexts and practices related to Bell's theorem, I hope to show that factors in addition to theoretical innovations, experiments, and techniques were necessary for the flourishing of research on this issue, and that the experimental implications of Bell's theorem were neither suddenly recognized nor quickly highly regarded by physicists. Indeed, I will argue that what was considered good physics after Aspect's 1982 experiments was once considered by many a philosophical matter instead of a scientific one, and that the path from philosophy to physics required a change in the physics community's attitude about the intellectual and professional status of the foundations of quantum mechanics. I have argued elsewhere (Freire, 2004) that a new attitude toward the foundations of quantum mechanics matured around 1970 related to subjects like the measurement problem and alternative interpretations of quantum mechanics, which were related neither to Bell's theorem nor to experimental tests. In the present paper, I argue that even concerning Bell's theorem and its tests a similar new attitude was required. On these events and periods there are already a number of testimonies, popular science books, and science studies works.[1] This paper, however, differs in that it attempts a historically oriented study about how what was considered a philosophical quarrel became a genuine topic of physics research.

Michael Horne, Abner Shimony, and Anton Zeilinger (1990) have produced a historical account of the concept of entanglement. These authors showed that as early as 1926 Erwin Schrödinger realized that this concept is a consequence of the mathematical structure of quantum mechanics, and that in the same year Werner Heisenberg explained the energy structure of the helium atom using states that are entangled. However, they also showed that in none of the first quantum mechanical treatments of many-body systems "was entanglement exhibited for a pair of particles which are spatially well separated over macroscopic distances" and that only with the Einstein-Podolsky-Rosen *gedanken* experiment, proposed in 1935, was this feature of the mathematical structure of quantum mechanics explicitly discussed. Finally, they showed that Schrödinger not only reacted to this ideal experiment by introducing the term "entanglement", but also asked himself if this quantum feature would be confirmed by experiments, or not. The authors continued sketching the historical record, passing through the appearance of Bell's theorem and its first tests until the appearance of down-conversion pairs of photons, in the late 1980s, led to improved tests of Bell's theorem. This paper has a narrower timeline. I focus on the period between 1965, when John S. Bell (1928-1990) published what we now call Bell's theorem, and 1982, when Alain Aspect published the results of his experiments violating Bell's inequalities and supporting quantum mechanics.[2] I leave aside Einstein's and Bohr's previous works and the debates on the interpretation and foundations of quantum mechanics in the 1930s, the debates on hidden variables triggered by the appearance of David Bohm's causal interpretation in the 1950s, and the ongoing series of new experiments on Bell's inequalities, since the late 1980s. Of these excluded topics, we only need to consider the context of the debates around Bohm's interpretation, since it strongly influenced the production and the initial reception of Bell's work. Indeed, Bell's decision to approach the hidden variable issue came from the very existence of Bohm's interpretation. Bell and his associates also inherited from the 1950s what, retrospectively, John Clauser (2003, p. 20) named the "stigma […] against any associated discussion of the notion of hidden variables in quantum mechanics."

---

[1] Aczel (2002), Bernstein (1981), Brush (1980), Clauser (2002a, 2003), Harvey (1980 and 1981), Pinch (1977), and Wick (1995).

The period in focus also allows us to discuss why optics became the privileged bench for experimental tests of Bell's inequalities. John Bell himself did not think it would be so at the beginning. It was those who first pushed these inequalities into the laboratories, such as Abner Shimony and John Clauser, who realized the conceptual advantages of optical tests when compared to tests with positronium annihilation, proton scattering, and other experiments. However, in addition to these advantages, other reasons operated in favor of optics. Training in optics was an asset of many who were willing to work on Bell's theorem and, for this reason, they could do their best and achieve telling results. By 1969 there was a balanced distributions of scientific skills, Clauser and Richard Holt being the optics experimenters and Shimony and Michael Horne the theoreticians without training in optics. After, a meaningful number of protagonists were trained in optics. One would like to inquiry further more about the connection between quantum optics and foundations of quantum mechanics. This question is not central to this paper, but I remark that Joan Bromberg (2005), who is working on the history of quantum optics in the U.S., suggested that while it is true that it has been useful for weaponry, it is also true that in this field "device physics was pursued in tandem with fundamental physics, and even with research into the foundations of quantum mechanics," and that device research led to fundamental physics problems, and the latter in turn inspired new devices. Indeed, technical improvements made available while the experiments were being carried out, such as the tunable laser, dramatically improved the accuracy of the experimental results.

---

[2] Bell (1964) was indeed published in 1965. Bell (1964) and Bell (1966) were written in the inverse order of their publication.

While following the theoretical, experimental, and technical issues related to Bell's theorem and its tests, I will pay attention to the biographical sketches of a few physicists involved in this story, addressing questions like: What factors led them to choose issues from the foundations of quantum mechanics as research themes? What issues did each one come to grips with? What were the favorable factors, and the obstacles, to their activities? To what extent did they succeed in their endeavor? Comparing their biographies, and thereby drawing resources from the method of prosopography in history and history of science,[3] I can draw a rough collective biography of these figures. As a whole, it is a story of success since they pushed a subject from the margins of physics to its mainstream; but it also included failures and hopes not fulfilled. It suggests to us that foundations of quantum mechanics was, at least in the period under analysis, a job for quantum dissenters, fighting against the dominant attitude among physicists according to which foundational issues in quantum mechanics were already solved by the founding fathers of the discipline. However, the common ground of the quantum dissenters was minimal and focused just on relevance of the research in the foundations of quantum mechanics, since these scientists supported different interpretations of this physical theory and chose different approaches and issues in their research. The different features of these biographies are as enlightening as their common traits are, and I will address both the contrasts and common features in this paper.

---

[3] For prosopographical methods, Stone (1971).

The first section of this paper deals with the context in which Bell's theorem was produced, its content, and its initial and uncomprehending reception; it covers the period between 1965 and 1969. Next I move on to analyze John Clauser and Abner Shimony's reactions, their proposal to submit the theorem to a viable experimental test, the involvement of American teams with these experiments, and the conflicting results among the first two experiments with optical photons. These events took place roughly between 1969 and 1974. The third section analyzes how the physicists involved settled the experimental tie with two new experiments carried out by Clauser and Edward Fry. This section is focused on the period between 1975 and 1976. It also treats the socialization of the physicists involved with research on Bell's theorem as well as the professional recognition of such themes of research. It pays attention to the cases of the journal *Epistemological Letters* and of the Erice 1976 meeting, which was seen by some physicists as the turning point in the acceptance that quantum nonlocality was indeed a new physical effect. The fourth section analyzes the road toward what was considered at the time the new challenge, an experiment changing the analyzers while the photons are in flight. It roughly deals with the period between 1976 and 1982, when Alain Aspect announced the results of his last experiments. Aspect's experiments resonated with the shifting attitude among physicists in the direction of a wide recognition of the importance of Bell's theorem. I conclude the paper by drawing conclusions on the way physicists changed their views on such subjects of research and by drafting a collective biographical sketch of the physicists who brought hidden variables to the optics laboratories.

**Bell's theorem, the context of its production, and its initial reception.**

The title of Bell's first paper, "On the problem of hidden variables in quantum mechanics," suggests a strong relation with the research program aiming to introduce hidden variables into quantum theory that was conducted by David Bohm, Louis de Broglie, and Jean-Pierre Vigier in the 1950s. However, the title cannot be taken at face value. Although they were historically intertwined, Bell's contributions did not give a new breath to this program. Rather, Bell led the hidden variable issue in a completely new direction. Indeed, Bell's approach and main achievements in quantum mechanics are of a very different strain when compared to Bohm's (1952) hidden variable interpretation. While Bohm built models that would first mimic quantum mechanics and later on lead to distinctive results, Bell was interested in the critical analysis of the assumptions behind mathematical proofs and *gedanken* experiments. This way, Michael Stöltzner (2002) has argued that despite Bell's enduring criticisms of von Neumann's proof of the impossibility of hidden variable in quantum mechanics, "a mathematically-minded view on the relation between the theorems of von Neumann and Bell" should consider Bell's theorem to be "a generalization of von Neumann's." Stöltzner's point is that "if one considers […] Hilbert's axiomatic method as a critical enterprise, Bell's theorem improves von Neumann's by defining a more appropriate notion of 'hidden variable' that permits one to include Bohm's interpretation which recovers the predictive content of quantum mechanics." However, Bell's work has a close historical connection with Bohm's work on hidden variable interpretation. He was directly motivated by the very existence of Bohm's proposal and by its reception among physicists. Bell's (1982 and 1987) statements – "In 1952 I saw the impossible done," and "Bohm's 1952 papers on quantum mechanics were for me a revelation" - hide more truth than is usually recognized.

Born in 1928, in Belfast, Bell had no scientific or educational family background; indeed, he was the first of his family to go to high school. He went to Queens University in Belfast, where he earned a BSc in Physics and formed the conviction that he would be a theoretical physicist. A job at the Atomic Energy Research Establishment at Harwell permitted him a leave of absence to begin his doctorate in Birmingham under Rudolf Peierls. Bell built his reputation working on high energy particle physics theory and the design of particle accelerators, and from 1960, he and Mary Bell, his wife, worked at CERN in Geneva. His early concerns about quantum mechanics can be traced back to his undergraduate courses in Belfast, where he quarreled with Richard Sloane, his teacher, because Sloane was not able to afford him a plausible explanation of Heisenberg's uncertainty principle (Whitaker, 2002, pp. 14-17). Later on, he avowed: "When I was a student I had much difficulty with quantum mechanics. It was comforting to find that even Einstein had had such difficulties for a long time" (Bell, 1982, p. 989). Since then he began to think about the transition between quantum and classical descriptions of the world.[4]

---

[4] We can reconstruct this account due to Bell (1982) and Jeremy Bernstein, who wrote his *Quantum Profiles* based on extensive talks with Bell and John Wheeler. Besides Bernstein (1991), biographical information on Bell can also be collected from Shimony (2002a), Whitaker (2002), and from papers by Bernard d'Espagnat, Michael Horne, and others, gathered in Bertlmann and Zeilinger (2002). For a comprehensive evaluation of Bell's scientific contributions, see Jackiw and Shimony (2002).

"Smitten by Bohm's papers,"[5] the Irish physicist attempted to determine what was wrong with von Neumann's proof, since it did not allow for hidden variables in quantum mechanics. Bell knew von Neumann's proof only indirectly, from his reading of Max Born's *Natural Philosophy of Cause and Chance*, but he could not read von Neumann's book because at that time there was no English edition of it. The solution was to ask Franz Mandl, his colleague at Harwell, about the content of the book. "Franz was of German origin, so he told me something of what von Neumann was saying. I already felt that I saw what von Neumann's unreasonable axiom was." He wrote to Pauli (1999, p. 28) asking for reprints of his paper on Bohm's proposal, but he probably did not like the views expressed there since Pauli (1953, p. 33) had considered Bohm's hidden variables as "artificial metaphysics." He went to Birmingham in 1953, including hidden variables as one of the possibilities for his studies. Asked by Peierls to give a talk about what he was working on, "Bell gave Peierls a choice of two topics: the foundations of quantum theory or accelerators." Peierls chose the latter, which was the end of the first stage of Bell's involvement with hidden variables. The intermezzo lasted ten years; he only resumed this work at Stanford, during a leave of absence from CERN. In the first of the two articles on foundations of quantum mechanics he published while in the U.S., the acknowledgements (Bell, 1966) record both the very origin of his investigation and former and latter influences: "The first ideas of this paper were conceived in 1952. I warmly thank Dr. F. Mandl for intensive discussion at that time. I am indebted to many others since then, and latterly, and very especially, to Professor J. M. Jauch." Discussions with Jauch had represented a true challenge for Bell, since Jauch "was actually trying to strengthen von Neumann's infamous theorem." According to Bell: "For me, that was like a red light to a bull. So I wanted to show that Jauch was wrong." In fact, Bell (1966) opened his first paper addressing it directly to Josef Jauch: "The present paper […] is addressed to those who […] believe that 'the question concerning the existence of such hidden variables received an early and rather decisive answer in the form of von Neumann's proof on the mathematical impossibility of such variables in quantum theory'" (p.447).

Bell's work can therefore be placed in the crossroad between the tradition related to the reinforcing of proofs against hidden variables and the tradition of building of viable models with such variables. If the possibility of introducing hidden variables in quantum mechanics was the motivation, Bell's approach was far from Bohm's, as we have remarked.

---

[5] Quotations are, unless indicated, from Bernstein (1991, 65-68).

Indeed, he was not interested in building viable models mimicking quantum mechanics. Instead, his works focused on the critical analysis of the assumptions behind von Neumann's proofs and its reformulations, and later on the assumptions behind the Einstein-Podolsky-Rosen *gedanken* experiment.

From the 1950s Bell received a heritage with a double meaning. In addition to the motivation derived from the existence of Bohm's work, he needed to face a widely shared idea that the matter of hidden variables was just a philosophical controversy and not a job for professional physicists. I have analyzed elsewhere the context of the disputes about Bohm's interpretation in the 1950s (Freire, 2005a). Here I just need to summarize it, pointing out that the label of "philosophical controversy" resulted from the overlapping of several factors. It originated with the physicists closely associated with the Copenhagen interpretation such as Pauli (1953, p. 33), Rosenfeld (1953, p. 56), and Heisenberg (1958, p. 131). "Artificial metaphysics," "debate [in] the field of epistemology," and "ideological superstructure," were respectively the words used to dismiss Bohm's hidden variable interpretation. These terms were used in the context that Max Jammer (1974, p. 250) named "the almost unchallenged monocracy of the Copenhagen school in the philosophy of quantum mechanics," describing the early 1950s. Bohm and collaborators unintentionally reinforced this label because the results they were able to obtain did not conflict with the predictions of quantum mechanics; neither did they present a heuristic advantage over that theory. In addition, Bohm, Vigier, and de Broglie emphasized what they considered to be epistemological advantages of their approach when compared to the complementarity interpretation. That the idea this was a philosophical controversy was largely shared can be seen from statements written by physicists who tried impartially to represent the controversy. Albert Messiah (1964, p. 48), in his very influential textbook, published originally in 1958, wrote, "the controversy has finally reached a point where it can no longer be decided by any further experimental observations; it henceforth belongs to the philosophy of science rather than to the domain of physical science proper." A similar example is Fritz Bopp's statement (Körner, 1957, p. 51), during a conference dedicated in 1957 to foundational problems in quantum mechanics: "…what we have done today was predicting the possible development of physics – we were not doing physics but metaphysics." Bopp's declaration becomes still more meaningful if one considers that he was working on another alternative interpretation, the so-called "stochastic interpretation." Finally, these widely shared views were reinforced by a pre-existent belief that foundational problems were already solved by the founding fathers of the quantum theory. In the middle of the 1960s the spirit of the "old battles" from the 1950s was still alive.[6]

Bell had a fine sensitivity for the prejudices nurtured in the old battles. He knew how strong the shared belief that he needed to face was. In the very period when he published his two seminal papers, he also published a third on the measurement problem in quantum mechanics, co-authored with Michael Nauenberg, in which they criticized the view shared by the *majority* of the physicists. "We emphasize not only that our view [that quantum mechanics is, at the best, incomplete] is that of a minority but also that current interest in such questions is small. The typical physicist feels that they have long been answered, and that he will fully understand just how if ever he can spare twenty minutes to think about it." As the paper was dedicated to Victor Weisskopf, they rhetorically appealed for Weisskopf scientific authority to support their own research: "It is a pleasure for us to dedicate the paper to Professor Weisskopf, for whom intense interest in the latest developments of detail has not dulled concerns with fundamentals." Ten years later, even with experiments on his inequalities underway, Bell kept the same sensitivity. When Alain Aspect formulated his proposal of new experiments on Bell's inequalities, he met Bell to discuss them. Bell's first question was, "Have you a permanent position?" After Aspect's positive answer, he warmly encouraged and urged him to publish the idea, but warned him that all this was considered by a majority of physicists as a subject for crackpots.[7]

Let us now focus on the results that Bell achieved in the middle of the 1960s. In his first paper, Bell (1966) isolated the assumption of von Neumann's proof that seemed to him to be untenable, and showed that while quantum mechanics satisfies this assumption it is not reasonable to require that any alternative theory have the same property. This assumption was that "any real linear combination of any two Hermitian operators represents an observable, and the same linear combination of expectation values is the expectation value of the combination." Next, he moved to analyze the new version of the proof that had been suggested by Jauch and Piron (1963), and made a similar objection. These authors had drawn an analogy between the structure of quantum mechanics and the calculus of propositions in ordinary logic, and they introduced a generalized probability function $w(a)$. They assumed as an axiom that "if $a$ and $b$ are two propositions, such that for a certain state $w(a) = w(b) = 1$, then this means that a measurement of $a$ and $b$ will give with certainty the values 1." Bell pointed out that this property, which is valid in ordinary logic and satisfied by ordinary quantum mechanics, should not be required of theories that were supposed to be alternatives to quantum mechanics. Bell also criticized Andrew Gleason's work (1957) on similar grounds for his use of quantum mechanical properties that were not reasonable to require of alternative theories. Bell remarked, however, that Gleason's work did not intend to reinforce proofs against hidden variables but rather to reduce the axiomatic basis of quantum mechanics.[8]

---

[6] "Old battles" was the term used by Abdul Salam in 1966 to explain to Rosenfeld an affair that happened at the International Centre for Theoretical Physics, in Trieste, Italy, related to a paper criticizing some tenets of the Copenhagen interpretation that had been written by the Brazilian physicist Klaus Tausk. Abdul Salam to Léon Rosenfeld, 26 Sep 1966, Rosenfeld Papers, Niels Bohr Archive, Copenhagen. Tausk criticized the "thermodynamic amplification" approach to the measurement problem, which had been proposed by the Italian physicists Daneri, Loinger, and Prosperi, and praised by Rosenfeld. Bell knew the Tausk affair. The case is described in Pessoa, Freire, and De Greiff (2005).
[7] Aspect (2002, p. 119). Michael Nauenberg (personal communication, 16 Apr 2005) also heard this anecdote from Bell.

After showing that previous proofs against hidden variables included assumptions that were not reasonable, Bell (1966) considered whether some features should be required from models with hidden variables, if these models were to be physically interesting. "The hidden variables should surely have some spatial significance and should evolve in time according to prescribed laws." He recognized that "these are prejudices," but added "it is just this possibility of interpolating some (preferably causal) space time picture, between preparation of and measurements on states, that makes the quest for hidden variables interesting to the unsophisticated." As the ideas of space, time, and causality had not been relevant in the assumptions hitherto considered, he attempted to determine what implications follow from hidden variables related to such ideas. After recalling that Bohm's 1952 proposal was "the most successful attempt in that direction," he wrote the wave function of a hidden-variable model for the case of a system with two spin ½ particles. Then, he showed that this wave function is in general not factorable and presents a grossly non-local character, since "in this theory an explicit causal mechanism exists whereby the disposition of one piece of apparatus affects the results obtained with a distant piece." As the state of two spin ½ particles could represent a system similar to that suggested by Einstein, Podolsky, and Rosen, Bell concluded, "in fact the Einstein-Podolsky-Rosen paradox is resolved in the way which Einstein would have liked least." Bell asked himself if non-locality is the price to be paid for the existence of hidden-variable theories, and admitted that there was no proof of this. Indeed, he was already looking for such a proof while writing the paper.

---

[8] It was Jauch who had called Bell's attention to Gleason's paper, which is an additional piece of evidence of how influential Jauch was in Bell's resuming of his own work on hidden variables.

To obtain such a proof, Bell took the next logical step: to isolate what reasonable assumption was behind Einstein's argument and check the compatibility between this assumption and quantum mechanics. For Bell (1964), the "vital assumption" when dealing with a two-particle system is that what is being measured on one of them does not affect the other. He recalled Einstein's dictum, according to which, "on one supposition we should, in my opinion, absolutely hold fast: the real factual situation of the system $S_2$ is independent of what is done with the system $S_1$, which is spatially separated from the former." Bell also knew that Bohm's hidden variable theory did not satisfy this dictum; so he went to build a simple model of a hidden variable theory obeying such a supposition and showed that its results conflict with quantum mechanical predictions in very special cases. This is Bell's theorem: no local hidden variable theory can recover all quantum mechanical predictions. In a very rough description, Bell's theorem can be derived when one considers a hidden variable model of a system with two spin ½ particles in the singlet state moving in opposite directions, a system that is analogous to the system suggested in the Einstein-Podolsky-Rosen argument. Bell built a function that is the expectation value of the product of spin components of each particle, and using different spin components derived an inequality with this function. The theorem is demonstrated when one uses quantum mechanical predictions in such inequality, since quantum mechanical predictions violate this inequality. Since then, many other analogous inequalities have been obtained, adopting somewhat different premises, as we will see along this paper; thus today it is usual to speak of Bell's inequalities as the quantitative measurement of Bell's theorem.

In spite of its simplicity, Bell's theorem has been considered by many one of the most important results in quantum physics since its creation in the middle of the 1920s, but awareness that the issue at stake was locality and not just hidden variables spread through a slow process. Even considering those who were already involved with foundational issues, not all of them quickly grasped the real meaning of Bell's theorem; what will be illustrated with David Bohm's and Louis de Broglie's cases. Research on the connections between quantum non-locality and relativity theory only began in the late 1970s, when the balance between experiments suggested confirmation of quantum mechanical predictions and violations of Bell's inequalities. How important as the cognitive obstacles were, however, they were not independent of attitudinal obstacles, which were mainly related to the intelectual and professional status physicists attributed to subjects such as hidden variables and foundations of quantum theory.

Simplicity of Bell's theorem has posed the following question, suggested by Shimony (2002): why was it Bell who arrived at such an "elegant but not very difficult" result? Reviewing Bell's steps in his career as a physicist, Shimony suggested that "Bell's moral character is primarily responsible for his discovery of Bell's Theorem," relating this discovery to his independence and tenacity in pushing critical analysis to its last consequences. As we will see, Bell actively participated in the endeavor to bring his theorem to laboratories, in the 1970s, and followed closely the wide recognition of his contribution in the 1980s. He died prematurely in 1990. In the next decade, Bell's theorem was the key concept behind the search for technological applications of quantum effects in quantum computation. Since then, his fame has only increased, and we may even be witnessing the birth of a new founder myth. According to physicist Daniel Greenberger (2002, p. 281), "John Bell's status in our field has the same [like Isaac Newton, James Watson, and Linus Pauling] mythic quality. Before him, there was nothing, only the philosophical disputes between famous old men. He showed that the field contained physics, experimental physics, and nothing has been the same since."

Let us now come back to the reception of Bell's theorem.[9] It opened the possibility of using experimental physics in order to reject some theories and preserve others; however,

---

[9] Bell's theorem paper was cited more than 2,000 times, 144 of them until 1982. I include information related to the number of citations of the main papers concerning the tests of Bell's theorem as evidence of their resonance among physicists. I am aware of the limits of such kind of information (Freitas and Freire, 2003), but it could help us just as one more piece of information. According to Igor Podlubny (2005), there is no reasonable criterion in the available literature for comparisons between "scientists working in different fields of science on the basis of their citation numbers." However, just as a guess, I took data from the number of U.S. article output and citations of U.S. articles, in the field of physics, for the years 1997, 1999, and 2000, available at www.nsf.org, and I obtained an average of 7.1 citations by article. Redner (2005), considering only citations in *Physical Review* of papers published in this journal, concluded that "nearly 70% of all PR articles have been cited fewer than 10 times" and that "the average number of citations is 8.8." These numbers match the physicists' shared tacit perception that an article should receive more than 10 citations to be

physicists did not see the subject in this way so promptly. Leslie Ballentine (1987) remarked, "the awareness of its significance was slow to develop," quoting a graph with the number of citations of Bell's 1964 paper. Indeed, consulting the *Web of Science* database one can check that before Clauser's (1969) note at the American Physical Society, and the Clauser, Horne, Shimony, and Holt 1969 paper, only three papers cited Bell's paper: Bell (1966), himself, and two letters that missed the point as concerns Bell's theorem.[10] Another evidence of the poor initial reaction to Bell's theorem can be found in the colloquium "Quantum Theory and Beyond" held at Cambridge in July 1969, which "intended to provide opportunity […] to discuss some possible alternative theories to see what a real change might involve." The colloquium was organized by Edward Bastin and David Bohm, chaired by Otto Frisch, and gathered physicists interested in foundations of quantum mechanics, such as Yakir Aharonov, Jeffrey Bub, Mario Bunge, H. J. Groenewold, Basil Hiley, Aage Petersen, G. M. Prosperi, and C. F. von Weizsäcker. None of them cited Bell's works.[11] Before going to Clauser's and Shimony's reactions, however, it is interesting to see that no reaction to Bell's theorem came from exactly where it would be expected, i.e. the partisans of hidden-variable approaches such as David Bohm and Louis de Broglie.

It is certain that David Bohm read Bell's 1966 paper, which contains references to the other paper in which the theorem was demonstrated.[12] Bohm could have exploited the full

---

known. *Spires*, Stanford's database for high energy physics preprints (www.slac.stanford.edu/spires/), suggests the following classification: "Unknown papers (0); Less known papers (1-9); Known papers (10-49), Well-known papers (50-99), Famous papers (100-499 cites), and Renowned papers (500+ cites)." Comparison with research on the foundations of quantum theory should be taken with a grain of salt due the huge difference in the number of active physicists in such fields. The source of data is the *Web of Science*, and for each paper I give first the whole number of citations and next the number up to 1982, the year in which Alain Aspect published his last experiment. The data were collected on 01 June 2005. I am grateful to Harvard University for the access to the *Web of Science*.
[10] Sachs (1969) cites Bell's paper just incidentally. Clark and Turner (1968) realized that Bell's theorem predicts a conflict between quantum mechanics and hidden variables, but they exploited neither the nature of this conflict nor viable tests to reveal this conflict.
[11] Bastin (1971). Henry Stapp claimed that a paper by himself, "widely circulated in 1968," was the first recognition of the importance of Bell's theorem. This paper "was to appear in the proceedings of Bastin's conference on Quantum Theory and Beyond, which occurred in the summer of 1968." Stapp to Clauser, 5 Feb 1975, Clauser Papers. I have not been able to unearth this unpublished paper. Indeed, these proceedings did not list Stapp among the attendance of the colloquium. The first paper published by Stapp, in which Bell's theorem is explicitly considered is Stapp (1971).
[12] Bohm & Bub (1966) cites Bell (1966), which cites the paper where the theorem is demonstrated Bell (1964).

implications of Bell's theorem. Instead, he, with his former student Jeffrey Bub (1966), reacted by building another type of hidden-variable theory, an explicitly nonlocal one. This time they used some ideas implicit in the "differential-space" theory of Norbert Wiener and Armand Siegel, and suggested a threshold, a relaxation time of the order of $10^{-13}$ sec, below which there would appear conflicts with quantum mechanical predictions.[13] They presented this theory as a candidate for solving the so-called measurement problem of quantum mechanics. As far as I know this was the only time Bohm suggested a figure to contrast hidden variables with quantum mechanics. In the 1950s, he had just begun to speculate that changes in his model could produce different predictions in the domain of the size of an atomic nucleus, but did not carry out the promised changes. Immediately after Bohm and Bub's proposal, the Harvard experimentalist Costas Papaliolios (1931-2002) tested it. Papaliolios (1967) successively measured linear polarization of photons emitted from a tungsten-ribbon filament lamp. The measurements were carried out within time intervals lesser than the threshold suggested by Bohm and Bub, and he found their theory untenable. Bohm was notified of the result before its publication and tried to reduce the reach of this experiment, "I regard our 'theory' largely as something that is useful for refuting von Neumann's proof that there are no hidden variables. I would not regard it as a definitive theory, on which predictions of experimental results could be made." In addition, he admitted that "the time, $\tau \approx \hbar/kT \approx 10^{-13}$ sec is just a guess," and not a consequence of their theory.[14] Papaliolios conceded that "the primary purpose of [Bohm-Bub's] paper was to demonstrate, by means of an explicit theory, how one can circumvent Von Neumann's proof," but emphasized the role that experimental predictions and real experiments, such as the one he carried out, should play in the choice between theories in the foundations of quantum physics.

[13] Previously, Wiener had been influenced by Bohm's works: "I have been tremendously influenced in my thinking by my conversations and correspondence with Mr. Gabor and Mr. Rothstein, and by reading a sequence of two papers […] which appeared this January under the authorship of David Bohm." In addition, in a talk given at the MIT-Harvard physics seminars, in 1956, Siegel cogitated of experimental predictions different from the standard ones, and N. F. Ramsey (Harvard) and Martin Deutsch (MIT), who attended the talk, "were quite willing to discuss the question as a serious and legitimate claim." "Paper to be presented on May 3 [1952] before the American Physical Society by Norbert Wiener," [7pp, unpublished, Box 29C, folder 678] and Armand Siegel to Norbert Wiener, 18 May 1956, [Box 15, folder 217], Norbert Wiener Papers. Wiener's interest in the foundations of quantum mechanics has not been analyzed yet, as far as I know, in the historical and philosophical literature on this subject.
[14] Costas Papaliolios to David Bohm, 17 Feb 1967; Bohm to Papaliolios, 1 Mar 1967, 2 Mar 1967, 11 My 1967. Papaliolios Papers, Boxes 23, folder "Hidden variables," and 10, folder "Bohm letters," respectively.

Papaliolios did not profess an empiricist view on choice of theories, he simply noted the advantages when experimental results are available. His reply to Bohm was a premonition of the coming times in the foundations of quantum mechanics:

> "It is to your credit that you make your theory testable by stipulating a definite relaxation time […]. If [it] had been left unspecified then you could always hide behind a suitably short relaxation time thereby making the hidden variable theory experimentally indistinguishable from the usual quantum mechanics. This latter approach is not entirely without merit, but one would have to use a non-experimental criterion such as elegance, simplicity, etc., in order to choose between the two theories." (Papaliolios to Bohm, 20 Mar 1967).[15]

While the Papaliolios' experiment did not deal with Bell's theorem, it is an interesting case for our study since it was the first time ever that an experiment was devised to test hidden variable theories. In 1957, Bohm and Yakir Aharonov had compared the ideal experiment suggested by Einstein, Podolsky, and Rosen with real data, but using results from a previous experiment that had not been designed with this goal. The context of Papaliolios' experiment indicates the shifting mentality concerning experiments and foundations of quantum physics among the physicists. It exhibits both old and new behaviors. Papaliolios was a Bell's theorem experimentalist *avant la lettre*, since for him Bohm-Bub's theory simply triggered what he had already been reflecting on, that is, he had been wondering about "hidden variable experimental possibilities," including a number of different possible approaches. For instance, he asked himself if "is there some effect which = 0 for quantum mechanics that $\neq 0$ for hidden variable theories?"[16] Aware of the old behavior of bias against the hidden variable subject, the U.S. Naval Ordnance Lab physicist Thomas Phipps praised the experiment and encouraged him to pursue this kind of experiments, noting, "even though this may not represent the most fashionable mode of research of the day."[17] Papaliolios received a similar and stronger encouragement from the *Physical Review Letters'* referee who evaluated his paper. "This paper should be published, but only if the author includes a discussion of the feasibility of these improvements and indicates plans to pursue the matter further." Papaliolios changed his paper according to the referee's requirement - "an experiment is now in progress to set even lower, upper bound on τ by using a thinner polarizer" -, and in fact elaborated plans to work out an improved experiment, but nothing came out of these attempts.[18] Yet, Papaliolios's optical experiment exhibits the interaction between scientific experiments and new technical devices that would be a driving force in Bell's theorem experiments. After his experiment, Papaliolios approached R. Clark Jones, Director of Research of Polaroid Corporation, showing how the polarizers produced by this corporation had been useful in a foundational experiment, and asking if Polaroid could supply thinner polarizers for new experiments.[19]

---

[15] Papaliolios Papers, box 23, folder "Hidden variables."
[16] "H. V. Experimental Possibilities?", minute by Costas Papaliolios, [w/d], Papaliolios Papers, box 23, folder "Hidden variables."
[17] Thomas Phipps to Papaliolios, 19 Apr 1967, Papaliolios Papers, ibid.

Louis de Broglie, the other main proponent of hidden variables, only reacted when experiments on Bell's theorem were already being carried out, and he did not grasp their full implications. De Broglie's argument, based on an analogy between quantum states and light wave packets, was that the wave function of a two-particle system would factorize after the particles fly a certain distance, which is a conjecture due to Wendell Furry. He did not realize that experimental tests of Bell's theorem could check this hypothesis. In addition, de Broglie (1974, p. 722) considered that quantum correlations between two electrons spatially separated would imply an instantaneous exchange of information, thus violating the relativity theory; an issue not yet elucidated. A harsh controversy followed between 1974 and 1978 that pitted Bell and Abner Shimony against de Broglie and George Lochak, a former student of de Broglie.[20] Lochak (1978) still maintained, "Bell's attempt, as interesting as it is, does not say, *and cannot say*, anything decisive about the existence of hidden variables, local or nonlocal." By that time, Bohm had realized the full implications of Bell's theorem, and stated that nonlocality was the most important quantum property.[21] In fact, Bell's theorem survived and was fully exploited in the hands of a new generation with very different approaches from those of the older one, as we will see now.

**Philosophy enters the labs: the first experiments**

The philosopher and physicist Abner Shimony and the physicist John Clauser were the key figures in the move to bring Bell's theorem to the laboratory. Before analyzing what did they produce let us see who they were at the time. Shimony, born in 1928, had been interested in science, mainly physics and mathematics, and philosophy since his undergraduate studies at Yale, where he was a joint philosophy and mathematics major.[22] When he was finishing his doctorate in Philosophy at Yale, working on probability, in the winter of 1952-1953, a reading of Max Born's *Natural Philosophy of Cause and Chance* revived his interests in physics, especially in classical statistical mechanics and quantum mechanics, and prompted him to take a second doctorate in physics under Eugene Wigner at Princeton. There he worked with statistical mechanics. In the early 1960s, even before finishing his physics doctorate, Wigner's interests in the measurement problem of quantum mechanics excited him.[23] Ever after, Shimony focused on the foundations of quantum mechanics, publishing his first paper on the subject in 1963. I have argued elsewhere that Wigner played a unique role in elevating the status of such issues. He was also very supportive of his students and colleagues who worked on this subject. His dispute with Léon Rosenfeld and the Italian physicists Daneri, Loinger, and Prosperi, in the second half of the 1960s, which contributed to breaking down the monocracy around the Copenhagen school, was primarily motivated by his defense of young physicists like Shimony who had been strongly and unfairly criticized. In the letter to Jauch suggesting him a joint reply to the Italian physicists, he wrote, "Needless to say, I am less concerned about myself than about other people who are much younger than I am and whose future careers such statements may hurt."[24] The lasting interaction with Wigner was fruitful for both of them. Shimony had Wigner's authoritative support for his entry into the field of foundations of quantum mechanics, and Wigner met in Shimony an informal assistant for philosophical matters.

---

[22] Shimony's biographical sketch is based on Wick (1995, 106-9), Bromberg (2004), Shimony (2002b), Aczel (2002, pp. 149-55), and Freire (in press).

[23] "I found your paper on the mind-body problem extremely stimulating. It is one of the few treatments of the problem which considers the mind-body relationship to be a legitimate subject for scientific investigation, without achieving this scientific status for the problem by reducing it to behavioristic or materialistic considerations." Abner Shimony to Eugene Wigner, 1 May 1961. Wigner Papers, box 94, folder 1. Shimony's first paper on the measurement problem is Shimony (1963).

[24] Eugene Wigner to Josef M. Jauch, 6 Sep 1966. Wigner Papers, box 94, folder 7.

Shimony's background in philosophy facilitated his physical research. Influence from Peirce and Whitehead facilitated his acceptance of quantum mechanics, and the remaining conflict, related to his commitment with realism, has been a major factor in his lasting "search for a world view that will accommodate our knowledge of microphysics." His double training also facilitated his professional career. He was hired by MIT's philosophy department in 1959, and gave courses on foundations of quantum mechanics there in the early 1960s. After a while, he moved to Boston University, in 1968, for a double affiliation in philosophy and physics. So he never depended exclusively on his physics training and his achievements in foundations of quantum mechanics for his professional career. His double training permitted him, however, to be considered a physicist with a philosophical culture among the physicists, and a philosopher with physics training among the philosophers. His achievements in the foundations of quantum physics have carried him to a key position in this field. With Shimony, foundations of quantum mechanics entered the optics laboratory, but did not lose its philosophical implications.[25]

John Clauser, born in 1942, has been uneasy about quantum mechanics since his undergraduate studies at Berkeley and graduate studies at Columbia, an uneasiness that he may have inherited from his family background. Francis Clauser, his father, was an aeronautical engineer and researcher who worked with von Karman on the physics of fluids. According to the younger Clauser, "he always was trying to understand physics, and there were very strong similarities between the mathematics of fluid flow and the mathematics of quantum mechanics, and he didn't understand quantum mechanics. And he kind of pre-programmed me as the guy who might help try to solve the problem that he couldn't solve." His father was also a strong influence on Clauser's skepticism, which was very instrumental for his discovery that no previous experimental data were adequate to test Bell's theorem. "Son, look at the data. People will have lots of fancy theories, but always go back to the original data and see if you come to the same conclusions."[26]

Clauser was finishing his thesis on measurement of the cosmic microwave background under the direction of Patrick Thaddeus at Columbia University when he became interested in Bell's theorem. Clauser's doctoral training, including the measurement of microwaves, enabled him to foresee and design quantum optics experiments to test Bell's inequalities. However, the appeal coming from the discovery of an interesting - but not yet done – experiment was not the only motivation behind his quick shift to a subject related to the foundations of quantum mechanics. Pedagogical and political factors were also influential factors. Clauser's approach to physics demands visualization and construction of physical models, not just abstract mathematics. So, he collided with the traditional way in which quantum mechanics has been taught.[27] Additionally, he (Clauser 2002a) read EPR's paper and Bohm's and de Broglie' works, and "while [he] had difficulty understanding the Copenhagen interpretation, the arguments by its critics seemed far more reasonable to [him] at that time." To his awareness that he had discovered that a good experiment had not yet been done, was added a political influence, "l'air du temps." As he remembers, "the Vietnam war dominated the political thoughts of my generation. Being a young student living in this era of revolutionary thinking, I naturally wanted to 'shake the world.' Since I already believed that hidden variables may indeed exist, I figured that this was obviously the crucial experiment for finally revealing their existence."[28] To shake quantum mechanics was then the target of his desire.

Since 1968, Shimony and Clauser, independently and without mutual knowledge, had been working on Bell's theorem. Indeed, Bell's paper early claimed Shimony's attention. "There is a paper by J. S. Bell […] which I found very impressive as evidence against hidden variable theories. He shows that in an Einstein-Podolsky-Rosen type of experiment the supposition of hidden variables, with any statistical distribution whatever, is certain to disagree with some of the predictions of quantum mechanics unless there is a kind of action at a distance;" he wrote to Eugene Wigner on New Year's day 1967.[29] In the summer of 1968, just before beginning to teach physics and philosophy at Boston University, Shimony enlisted the physics graduate student Michael Horne for designing a "realizable Bohm-type EPR experiment" as a dissertation subject. Horne was the right choice as he was attracted to physics for the intellectual endeavor it embodied. He was captivated by physics while reading I. B. Cohen's *The birth of a new physics* in high school, and the reading of E. Mach's *The*

*science of mechanics*, while at the University of Mississippi, led him to decide that he wanted to do research on the conceptual foundations of physics. In the early 1969, however, Shimony was surprised by Clauser's abstract in the *Bulletin of the American Physical Society* suggesting an experiment to test Bell's theorem. "We were scooped," told Shimony to Horne.[30] After consulting Wigner, Shimony decided to call Clauser and suggest collaboration, which was accepted by Clauser. By the time they began to collaborate they had already independently realized that no previously available experimental results were able to test Bell's theorem and that the most adequate and viable test would be to repeat in slightly different conditions an optical experiment done by Carl Kocher as a doctoral student of Eugene Commins in 1967. The CHSH paper, which will be analyzed later, was the first result of this collaboration.[31]

The first news Bell had from the American reaction to his work did not come, however, through Shimony. According to Wick (1995, p. 106), "The letter from the American student was the first serious reaction [Bell] got to his paper – after a lag time of five years." The American student was, surely, Clauser. It is worth noting the letters exchanged between Clauser and Bell, not only for what they correctly predicted but also for their unfulfilled hopes. Clauser rightly assured Bell that the results from the Wu-Shaknov experiment with the annihilation of positronium were not adequate to test Bell's inequalities. He suggested, instead, a modified extension of the Kocher and Commins experiment with the polarization correlation of photons from an atomic decay cascade. He also promised that "it might also be possible to 'rotate' the polarizers by means of magneto-optic effects while the photons are in flight to rule out all local hidden-variable theories;" a promise that would wait more than ten years to be fulfilled, and then not by Clauser but by Alain Aspect with a different method for obtaining time-varying analyzers.[32] Bell rightly anticipated the results of the experiments, holding slight hope for a breakthrough: "In view of the general success of quantum mechanics, it is very hard for me to doubt the outcome of such experiments. However, I would prefer these experiments, in which the crucial concepts are very directly tested, to have been done and the results on record. Moreover, there is always the slim chance of an unexpected result, which would shake the world." He also revealed that he expected to be performing experiments in particle physics: "experiments have been proposed involving neutral kaons […] and will become practical in the course of time."[33] Particle physics, though, never became the main bench for experiments with Bell's theorem.

---

[30] Horne (2002), and Horne (personal communication, 8 June 2005).
[31] For the roads of Clauser and Shimony to Bell's theorem, and for their meeting, see Wick (1995, pp. 103-113), and Aczel (2002, pp. 149-169).
[32] John Clauser to John Bell, 14 Feb 1969, Clauser Papers.

It is interesting that among the three quantum dissenters, Clauser and Bell were more optimistic about the possibility of obtaining results violating quantum mechanics than Shimony, and that Clauser was by far the most optimistic among them. Before they met each other, Shimony wrote to Clauser: "Incidentally, I am amazed at your estimate of the probabilities of the possible outcomes of the experiment. I would estimate a million to one in favor of the quantum mechanical correlation function. Needless to say, I hope I am wrong in this." "Do keep imagining that it will come out *against* quantum theory; that makes it very interesting!" were words from Horne to Clauser. Shimony kept these memories of Clauser's hopes, "… he was absolutely convinced that the experiment was going to come out for the local hidden variable theory and against quantum mechanics, and it was going to be an epoch-making experiment."[34]

The CHSH 1969 paper is a fine piece in physics literature both for its concision and breadth.[35] It is an acronym for Clauser, Horne, Shimony, and Richard Holt, a Harvard University student of Francis Marion Pipkin (1925-1992). The authors fulfilled three goals. They modified Bell's theorem in order to make it usable for realizable instead of ideal experiments, they showed that data available from previous experiments did not produce evidence against local hidden variable theories, and they suggested a viable test with optical photons suggesting optics as the privileged bench for tests of Bell's theorem.[36] They showed that the experiment by Wu and Shaknov, performed in 1950 using the annihilation of positronium for measuring polarization correlation of γ photons was not adequate to test Bell's theorem because "the direction of Compton scattering of a photon is a statistically weak index of its linear polarization," and there are no good polarizers for high energy photons. They also showed that the Kocher and Commins experiment measuring polarization correlation of photons emitted in an atomic decay cascade of calcium, performed in 1967, was in principle adequate for testing Bell's theorem but that these experimentalists had only measured angles - 0° and 90° - in which there is no conflict between quantum mechanical predictions and local hidden-variable theories. For this reason the proposed experiment was essentially a revision of the Kocher and Commins experiment with angles - 22.5° and 67.5° - in which there is a maximum conflict of predictions. It is curious to remark that Kocher and Commins (1967, p. 575) explicitly presented their experiment "as an example of a well-known problem in the quantum theory of measurement, first described by Einstein, Podolsky, and Rosen and elucidated by Bohr," but they were not aware of Bell's work. In order to make Bell's theorem testable, CHSH authors introduced additional assumptions that were determined by the type of experiments they wanted to do. The three additional assumptions were related to: avoiding Bell's assumption of perfect correlation between the pair of particles; considering rates of emergence or not of photons from the filters instead of their detection (a change due to the small efficiencies of the available photoelectric detectors); and taking the probability of joint detection as being independent of the orientation of the polarization filters. They considered that the later assumptions, which can be called a fair sampling assumption, "could be challenged by an advocate of hidden-variable theories in case the outcome of the proposed favors quantum mechanics;" but they did not assess them a flaw in the proposed experiment because "highly pathological detectors are required to convert hidden-variable emergence rates into quantum mechanical counting rates." As necessary as these assumptions were for the time we are analyzing, to perform experiments relaxing them has been the holly grail for physicists involved with foundations of quantum mechanics.

---

[33] Bell to Clauser, 5 Mar 1969. Clauser Papers.
[34] Shimony to Clauser, 20 Apr 1969, Horne to Clauser, 18 Apr 1969, Clauser Papers; Shimony (2002b, p. 71).
[35] The CHSH paper has 742 citations, 92 of them until 1982.
[36] This suggestion faced competition to be established. The competition pitted the teams involved with atomic cascade and those with positronium annihilation. When both experiments were already done, Shimony wrote to Clauser (19 May 1072, Clauser Papers): "Freedman told me about the difficulties raised by Wu, Ullman, and Kasday. What has been the upshot of that? I think the only way to handle it is to continue to state, politely but firmly, that their experiment is a fine one, but much less decisive than yours, because of their additional assumption."

The CHSH paper awakened interest among Europeans physicists, such as Bell, Bernard d'Espagnat, de Broglie and Franco Selleri, who were already involved with hidden variables theories. As a consequence of this paper, Shimony was invited by d'Espagnat to lecture at the Varenna school on foundations of quantum mechanics, in 1970. The Varenna summer school, that is, the "International School of Physics 'Enrico Fermi'," had been organized by the Italian Physics Society starting in 1953, and had become a traditional means for training European physicists in novel themes of research. The 1970 school was the first to be dedicated to the theme of the foundations of quantum mechanics. There, Shimony gave three lectures, made acquaintance with and began a lasting friendship with Bell and d'Espagnat, and became still more involved with issues related both to Bell's inequalities and measurement problem. We can conclude that by the early 1970s Shimony had received recognition in Europe and the United States for his work on foundations of quantum mechanics. It is also worth remarking that the Varenna school was an important meeting point for developing research on Bell's theorem as it brought together a number of physicists who were already working on the subject but had never met each other previously.[37]

From the CHSH authors emerged the two teams that conducted the first experiments with optical photons. At Berkeley there was Clauser, who moved there as a postdoc of Charles Townes, and Stuart Freedman, a doctoral student of Commins. At Harvard, there was Holt, a student of Pipkin. Holt's work involved not only Pipkin, his adviser, but also Costas Papaliolios, who two years before had performed the experiment on the Bohm–Bub theory. At Berkeley, Clauser had the support of Charles Townes for the experiment. "Townes was the guy who actually twisted Commins' arm to put Stu Freedman on the experiment and to steer Atomic Beam Group funds into doing the experiment." However, Clauser did not receive a wide support from the Berkeley faculty. Clauser remembers that even after the experiment was performed "most of the physics faculty at Berkeley all said [this was junk], because as 'you got exactly what you expected, what was the point?" For Clauser, 'they did not understand Bell's Theorem." Commins, who had performed the experiment that Clauser was repeating with some modifications, was no exception; "what a pointless waste of time all of that was" are Clauser's (2002b, p. 12-13) memories of Commins's remarks. In fact, as early as February 1969, Commins had "dismissed the idea [repetition of his experiment] as worthless," and stated that "there are 'thousands' of experiments which already prove" what Clauser was looking for, which led Clauser to appeal for Townes' mediation.[38] Harvard's environment was very different. In 1969, even before his meeting with Clauser, Shimony was enthusiastic about the involvement of Harvard experimentalists. "There is a great deal of interest in the experiment at Boston University and at Harvard, though I think it could be done quickly only at Harvard. [...] The man most interested at Harvard is Costas Papaliolios. [...] However, he is very busy now, and therefore suggested it to two men whom I have not met yet. One is Nussbaum, who did his doctoral works at Harvard under Pipkin, looking at photon polarization correlation in a mercury cascade. The other is Dick Holt, another student of Pipkin who inherited Nussbaum's apparatus."[39] In the late 1969, the possibility of three tests of quantum mechanics caught the attention of popular science magazines. "Acid test for quantum theory," advertised *Scientific Research*, explaining that "three versions of the same experiment, just getting under way at three separate laboratories, may supply a definitive answer as to whether there are indeed 'hidden variables' that provide a more deterministic of reality than is possible with quantum mechanics."[40] Gilbert Nussbaum, then at the Bell Laboratories, did not, in fact, comply, but his place was occupied later on by the Texas A&M University physicist Edward Fry, who would be the leader of the third team to carry out experiments with optical photons.

---

[37] Varenna's school had 84 participants. For its proceedings, d'Espagnat 1971. For its stimulus on the research on foundations of quantum mechanics, Freire (2004). Shimony (2002, pp. 75-82) thinks he was invited to Varenna for his previous work on measurement problem. However, d'Espagnat (2001) remembers that it was Bell who suggested he invite Clauser and Shimony.

[38] Clauser to Eugene Commins, 18 Feb 1969, [cc: C. Townes], Clauser to Townes 18 Feb 1969, Clauser Papers.

Fry had been trained in atomic physics and spectroscopy, while doing his thesis at the University of Michigan under Bill Williams. He was apointed as Assistant Professor at Texas A&M, in 1969. There he was introduced by James McGuire, a theorist in atomic collision physics, to the CHSH paper, in "evening philosophical society discussion in late fall 1969." Fry was "immediately intrigued," figured out an experimental scheme, and applied to National Science Foundation for funding it. It was not a good experience for his first ever application. "The reviews did not argue about the physics, their theme was basically that NSF should not waste money on fruitless pursuits." Discouraged by the result and by technical troubles with the available equipments, he did not actively pursue the intended experiments until after Clauser and Holt results. Meanwhile, his bets for the results of such experiments were against quantum mechanics predictions. He considers that he "had the same bug as Clauser and may have even, in part contracted it from him." According to his reminiscences, "I was really hoping for a major breakthrough in our understanding of the quantum world.

Clearly, a violation of the Bell inequalities does improve our understanding; but it does not provide the dramatic overthrow of existing thought that I had anticipated."[41]

The first round of experiments ended in a tie. The experiment conducted by Freedman and Clauser (1972) confirmed the quantum mechanical predictions and violated Bell's inequalities, while the experiment held by Holt (1973) under the advice of Pipkin produced the opposite results.[42] Freedman and Clauser observed pairs of photons emitted by transitions in calcium. They used "pile-of-plates" polarization analyzers, and the experiment ran for 200 hours. For a certain variable that resulted from count rates of photons they obtained $0.300 \pm 0.008$. For the sake of simplification, let us name this magnitude S. Bell's inequality for this variable and this experimental setup was $S \leq \frac{1}{4}$, and the quantum mechanical prediction for the same variable was $0.301 \pm 0.007$. Holt and Pipkin observed photon pairs from transitions in the isotope of mercury $^{198}$Hg. They used calcite prisms as polarization analyzers, and the experiment lasted 154.5 hours. For this case, the quantum mechanical prediction for the same variable was 0.266, and they obtained $0.216 \pm 0.013$. As we will see later, the main effect on the physicists was the desire for new experiments. Clauser repeated Holt's experiment, Fry used new techniques for getting better results, and Aspect devised what he intended to be an experiment able to settle the controversy.

In addition to these experiments with polarization correlation of optical photons there appeared experiments in fields other than optics. The experimentalist L. Kasday (1971), a doctoral student at Columbia, the Italian group of G. Faraci in Catania and the group led by A. Wilson at Birkbeck College, in London, repeated the Wu and Shaknov experiment with $\gamma$ photons, with further assumptions, and arrived at conflicting results. Kasday's and Wilson's results confirmed quantum mechanics while Faraci's results confirmed Bell's inequalities. At Saclay, France, M. Lamehi-Rachti and W. Mittig performed an experiment with spin correlation in proton-proton scattering, confirming quantum mechanics.[43] In the ten years

---

[41] Edward Fry (personal communication, 5 Aug 2005). James McGuire to Clauser, 3 Jan 1972, and 24 Feb 1972, Clauser Papers. The collaboration with James McGuire led to a derivation of Bell's theorem and comparison with data from the experiment by Freedman and Clauser (McGuire and Fry, 1973). For the technical troubles, see Harvey (1980, p.154).

[42] Freedman and Clauser's paper has 343 citations, 114 until 1982.

[43] Kasday (1971), and Kasday et al (1975). The latter paper has 28 citations, 13 until 1982. Faraci et al (1974) has 51 citations, 37 until 1982. Wilson et al (1976) has 43 citations, 22 until 1982. Lamehi-Rachti and Mittig (1976) has citations, 21 until 1982. After 1978,

between Freedman and Clauser's result in 1972 and the three experimental results Aspect and his collaborators would publish in 1981 and 1982, we had 11 experiments and 8 teams involved with tests of Bell's theorem.

Since experiments with optical photons were considered by Shimony, Clauser, and Bell the most adequate to test local hidden variables, and the resonance of results from atomic cascade experiments evidences that this stance was widely shared by the physicists involved with Bell's theorem, it is interesting to focus on the period between 1973 and 1974, when the only available results from such experiments were in conflict with each other, and see how the physicists reacted to this disagreement. The physicists involved in theses issues did not consider that there was a true tie. Harvard physicists did not trust their own results but suspected them of systematic errors, which, however, they could not identify them. They decided not to publish the results, but Holt's work was recognized and he received his PhD. Behind this decision was their trust in quantum mechanics. Pipkin concisely recorded that "the measurements on the polarization correlation disagree with quantum mechanics and agree with the predictions of hidden variable theory. So far efforts to explain this discrepancy have not been successful."[44] He did not ponder the possibility that their results had exposed a deficiency of quantum theory. Holt, in his thesis, dared more, writing, "the polarization correlation results present a far more puzzling (and perhaps) exciting prospect. The statistical accuracy is certainly great enough to allow us to say that a discrepancy with quantum mechanics exists. The arguments of the preceding section show that all the obvious sources of systematic errors have been examined." Next he attenuated his claim, "on the other side are two arguments for disbelieving the results," the first one being the results already published by Freedman and Clauser. He did not discuss why this previous result should be considered more reliable than his own results. Next, he stated the main reason for his caution, "the second argument against our results is the enormous success that has been enjoyed by quantum mechanics in the correct prediction of experimental results." Holt's (1973, p. V-27) awareness that this argument could not be definitive led him to conclude, "this, however, is not a telling argument because it is quite conceivable that a deterministic theory substructure could yield the same ensemble average as quantum mechanics in all other experimental situations save this one, in which the least intuitive features of quantum mechanics are strikingly displayed." Holt's back and forth reasoning did not surprise Bill Harvey (1980, p. 143), who has written sociological studies on these experiments. For him, "despite its possible philosophical significance, [local hidden variables] could hardly be described as a highly *plausible* theory." Yet, he showed how the idea of plausibility, widely used by the physicists involved in the debate on Bell's theorem, was strongly dependent on "their immersion in the *culture* of physics" and could not be reduced to logical reasoning or data evaluation Harvey (1981, p. 105). Trust in quantum mechanics was, and is, an essential part of the culture of physics.

We have good evidence of this trust in quantum theory in a review of the results written in 1974 by the French physicist Michel Paty. He was at the time an experimentalist in particle physics at Strasbourg, very interested in epistemological issues, preparing himself for a conversion to a philosophical career by taking a second doctorate, this time in philosophy. Together with the Brazilian physicist José Leite Lopes, he ran a series of seminars and a journal under the title *Fundamenta Scientiae*. In 1974, they organized a colloquium dedicated to the 50[th] anniversary of quantum mechanics in which Paty presented a review on "the recent attempts to verify quantum mechanics." After concluding that "the present balance sheet of the experiments designed to test Bell's inequalities is therefore as follows: three agree with quantum mechanics, and two disagree;" he asked, "has quantum mechanics now revealed its limitations, or more exactly, the limits of its field of application?" He conjectured that "This would not be unthinkable a priori, […]. This would also be the case for a theory as powerful as quantum mechanics, which itself is highly powerful, but at the same time probably has a frail basis." Next, however, he did not accept his own conjecture, and reaffirmed a trust in quantum mechanics: "However, it may seem doubtful that such an established theory might be questioned in such simple experiments. And in fact quantum mechanics may only *appear* to be frail; its hold on our conceptions is paradoxically shown in this recent questioning: it is not quantum mechanics which is put into doubt, so much as the basis of these very experiments or at least their interpretation." Anyway, Paty was careful about this conclusion, and urged for more refined experiments. [45] This kind of discussion shows the tacitly shared view that foundational problems in quantum mechanics were already solved by its founding fathers being reinforced by the increasing practical success of this theory. Even physicists like Pipkin, Holt, and Paty, who spent time doing or reviewing such experiments and for this reason cannot be counted among those who developed prejudices against hidden-variable theories as a subject, embodied this trust in quantum mechanical predictions.

---

[45] Paty (1976). He included in his count not only optical photons experiments, but also proton-proton scattering and positronium annihilation experiments.

Now I want to take the case of the Harvard experimenters for close consideration. Compared to Berkeley, Harvard seems to have been a friendly intellectual and professional environment for those who were interested in Bell's theorem. We can infer this both from the involvement of its experimenters with tests of hidden variables and from the testimony of Holt. Papaliolios was a key figure in the creation of this intellectual and professional environment, since he was openly involved with tests of hidden variables since 1967 and throughout the 1970s. Acknowledgements in Holt and Horne's dissertations, and in the CHSH paper testify to his role in the first tests of Bell's theorem. He published with Freedman and Holt a review on the status of hidden variable experiments, and in 1975 gave a course at Harvard on hidden variables.[46] Papaliolios earned a PhD from Harvard in 1965, and was a professor of physics there from that time until his retirement in 2001, while being also a physicist at the Harvard-Smithsonian Center for Astrophysics. His main research interests were related to astrophysics, but he included hidden variables as one of his topics of research in Harvard's reports to its visiting committees even before he was tenured as a Professor in 1971.[47] It was Papaliolios who invited Pipkin to undertake an experiment on Bell's theorem. Pipkin went to Harvard after his PhD in Physics at Princeton and there became a member of the faculty in 1957. Reputed as a good experimentalist both in low energy atomic physics and high-energy particle physics, Pipkin had begun in the middle of the 1960s a line of research on precision measurement of the atomic fine and hyperfine structure, which was continuously funded by the National Science Foundation. This approach included the study of correlation of photons from atomic cascade in order to calculate lifetime of atomic states. Papaliolios proposal was easily accommodated in Pipkin's project through the thesis work of Holt, his doctoral student, but Pipkin's interest in foundations of quantum mechanics were never as strong as Paliolios' were. Holt's history at Harvard presents nuances that show us distinctive features in the social and intellectual recognition of the importance of Bell's theorem.[48] In general, he did not encounter prejudices or disdain but merely a lack of interest. Edwin

---

[46] Freedman, Holt, and Papaliolios (1976). The readings for this course are in Papaliolios Papers, box 16, folder "Fall 75 Hidden Variables – Reading Course – (P351)."
[47] Reports of Visiting committees are in Costas Papaliolios Papers, box 5, folder "Visiting Committee, 1970-1973."
[48] Richard Holt (personal communication, 21 Mar 2005). See also Wick (1995, p. 108).

Purcell was on the committee examining his thesis and considered the subject "a worthwhile endeavor," but in his introductory quantum mechanics course, which Holt attended, quantum measurement was handled via Schwinger's measurement operator algebra, in which the full measurement problem is not explicit.[49] As a different example, he has the vivid image of the graduate quantum mechanics course given by Paul Martin, in which not only measurement was carefully presented via Kurt Gottfried's textbook, but also Bell's theorem was introduced.[50] Holt read Bell's paper but did not become interested himself, and considers that even if he was not yet a physicist he shared the attitude of the time. Thus, Holt does not speak of stigma associated with the subject but rather of little interest in the subject, and even this little interest was not unanimous. As a sign of the times, Holt (1973) opened his thesis with an approach that in other times and places would have been considered heretical. He analyzed the measurement problem in quantum mechanics and presented the "Copenhagen interpretation" as one of its possible solutions, the others being the "Everett-Wheeler interpretation," the "Wigner's idea," and the "hidden variables" (p. I-6-14). Incidentally, I remark that, fifteen years before, regarding the Copenhagen interpretation as only one of the possible interpretations of quantum mechanics was removed from a dissertation before its publication. It happened with Hugh Everett's dissertation at Princeton under John Wheeler.[51]

As favorable as Harvard was for the experimental tests of Bell's theorem, its participation in our account is not a story of success. Holt and Pipkin did not trust their result, but they were not able to identify the source of errors. They circulated it as a preprint that was never published, meaning that in fact they did not claim to have found a violation of quantum mechanics prediction. In his sociological studies Harvey discussed how their cautious stance was conditioned by local and cultural circumstances such as Holt's status as doctoral student, trust in quantum theory, and a previous failure of Pipkin's, who had a violation of quantum electrodynamics that was not eventually confirmed.[52] As important as these factors could have been, and were, I think one should add a fact that was not determined by such factors. Pipkin and Holt gave up the subject, they did not pursue it to its ultimate consequences, by

---

[49] The composition of the committee who examined Holt's thesis was Pipkin, Papaliolios, and Purcell.
[50] Gottfried (1966). The whole section IV is dedicated to "The measurement problem and the statistical interpretation of quantum mechanics." Bell's 1966 paper is suggested for reading.
[51] Everett (1957) and (1973). On Everett's case, Freire (2004, pp. 1750-53).
[52] Harvey (1980). Pipkin's claim of a violation of quantum electrodynamics was Blumenthal et al (1965).

repeating the same experiment or by planning a new one, in spite of the interest their unpublished result awakened even beyond the group of physicists already involved with Bell's theorem.[53] Thus, the University of Southern California physicist Marc Levenson wrote to Pipkin, "I have obtained a preprint of your paper with Holt which casts doubt upon the validity of quantum mechanics. This result distresses me somewhat as I am expected to introduce our juniors to this subject next semester," and Levenson continued discussing possible sources of error.[54] One can conjecture that neither Pipkin nor Holt were able to foresee the importance that this subject would acquire in physics. Pipkin only took it up again in 1976, when Clauser and Fry were announcing the new results we discuss below. Pipkin then stated, "a careful study was made of systematic effects which could account for the deviation from the quantum mechanical prediction but no candidates were found. In view of the result reported by Freedman and Clauser […] it was concluded that the experiment should be repeated with a somewhat different configuration of the apparatus."[55] However, they had not tried and they did not try any repetition; it was up to Clauser, Fry, and later Aspect to make new experiments on Bell's theorem. Two years later, Pipkin made his "closing arguments," while reviewing the "atomic physics tests of the basic concepts of quantum mechanics." After repeating that they had "recommended that the experiment be repeated by someone else with a different configuration of apparatus," he cited Clauser's new results and concluded, "this experiment thus indicated that the Holt-Pipkin experiment was incorrect although it did not localize the source of the error in the earlier experiment."[56] Their failure to analyze what was wrong with their experiment was probably responsible for the deletion of their role in the current story of success associated with Bell's theorem. So, Amir

---

[53] When Clauser was repeating Holt's experiment, the latter wrote to the former, "every time that Stu Freedman asks me when we're going to publish, I tell him I'm waiting for your results." Holt to Clauser, 31 Aug 1975, Clauser Papers.

[54] Marc Levenson to Pipkin, 03 Dec 1974. Pipkin Papers [Accession 12802], box 12, folder "NSF proposal 1974-1975."

[55] Holt and Pipkin (1976). The report was presented at the Erice workshop by Pipkin. Similar words appeared in Freedman, Holt, and Papaliolios (1976).

[56] Pipkin (1978, 317-9). This review received 33 citations. Until today, Holt holds the same opinion, "I think it is still accurate to say that the source of the error remains unknown." Richard Holt (personal communication, 21 Mar 2005). Clauser and Shimony's conjecture is that "stresses in the walls of the Pyrex bulb used to contain the electron gun and mercury vapor" made the glass optically active, and this systematic error was not adequately compensated. A similar problem appeared while Clauser was repeating the experiment. After the stresses were removed, "the experiment was re-performed, and excellent agreement with quantum mechanics was then obtained. On the other hand, Holt and Pipkin did not repeat their experiment when they discovered the stresses in their bulb." Clauser & Shimony (1978, p. 1910).

Aczel (2002), in his popular science book *Entanglement*, dedicated one chapter to "the dream of Clauser, Horne, and Shimony," and another to the "Alain Aspect." Holt, Pipkin, and even Edward Fry's roles were sent to the backstage of the history. Symptomatically, Pipkin's Harvard colleagues, while writing his official obituary, did not include experiments in foundations of quantum mechanics among his achievements.[57]

Describing Holt and Pipkin's case as a failure risks anachronism.[58] After all, experiments of Bell's theorem became mainstream physics in the 1980s, and one needs to consider the importance of the result for them, at the time they performed the experiment. Indeed, neither for Pipkin nor for Holt did this experiment have the importance that we attribute to it today. Pipkin was developing a line of research on precision measurement of the atomic fine and hyperfine structure, from which Nussbaum's thesis on lifetime of atomic states obtained through atomic cascade experiments was one of the first results (Nussbaum and Pipkin, 1967). Holt's experiment on Bell's theorem was a small extension of this method.[59] Indeed, in Pipkin's 1972 NSF proposal the completion of Holt's experiment is listed as #6 out of a list of seven goals, while it also included a precision measurement of the lifetime of the $7^3S_1$ state of atomic mercury, a result that was indeed published.[60] Apparently, Pipkin's reputation as experimenter was not damaged by the unsolved problem with the test of Bell's theorem. In 1990, the physicist who evaluated his NSF grant extension in the same domain stated, "Professor Pipkin is a well-established leader in the field of high-precision measurements of fundamental atomic systems," and rated the proposal as "excellent."[61] Holt made a successful career working with precision measurement in atomic physics at the University of Western Ontario, Canada, but never returned to the subject of Bell's theorem.

---

[57] Gary Feldman, Paul Horowitz, Costas Papaliolios, Richard Wilson, and Robert Pound (Chairman), "F. M. Pipkin – Memorial Minute," *Harvard Gazette*, 26 Nov 1993, p. 15. At Harvard, this kind of obituary is commissioned. See Jeremy Knowles to Papaliolios, 10 Apr 1992. Papaliolios Papers, box 26, folder "Frank."

[58] Harvey (1980, p. 158), spoke of "Holt's virtual capitulation." I think that he singled out too much Holt's profile as a graduate student. Indeed, as we have seen, it would be more reasonable to describe the case as a story of failure of the Harvard experimentalists involved, which was responsible for the deletion of their participation in the present story of the success of Bell's theorem.

[59] In his "Proposal: Atomic Physics Experiments Using Photon Coincidence Techniques," [1969], Pipkin listed as goal #2 "To continue the present coincidence measurements of the 4358Å – 2537Å photon cascade in mercury […]," and listed Holt as doctoral student, but no reference was made to the hidden variable test, which would only appear in the proposal of the next year. See F. M. Pipkin, "Atomic Physics Experiments Using Fast Atomic Beams and Photon Coincidence Techniques," NSF Grant Proposal [#GP22787], [1970]. He was funded $54,200.00 for two years, cf. Rolf Sinclair [NSF] to Pipkin, 26 May 1970. Pipkin Papers [Accession 12802], box 23, folder "NSF Atomic 1973."

[60] F. M . Pipkin, "Proposal for a Grant from the NSF to continue atomic physics experiments using fast atomic beams and photon coincidence techniques." He was funded $112,380.00 for two years. Rolf Sinclair [NSF] to Pipkin 25 May 1972. Pipkin Papers [Accession 12802], box 23, folder "NSF Atomic 1973." Holt and Pipkin (1974).

The 1970s was the decade when hidden variables, an issue once considered a question of philosophical taste, entered the lab. These experimental activities did not mean, however, a decline in theoretical work on hidden variables. There was a flow of new derivations of Bell's inequalities. In 1978, reviewing the subject, Clauser and Shimony analyzed at least eleven different derivations, by Bell himself, Wigner, Frederik Belinfante, Holt, Clauser and Horne, Henry Stapp, d'Espagnat, D. Gutkowski and G. Masotto, Selleri, and L. Schiavulli, in addition to a new and different derivation by Bell, which was criticized by Shimony, Horne, and Clauser.[62] Two new books appeared, one by Belinfante (1973), entirely dedicated to the hidden-variable issue, and another by d'Espagnat (1976), dealing with the foundations of quantum mechanics. The derivations made by Bell, Clauser, and Horne helped to focus what he issues at stake were in these experiments. Bell (1971) presented a proof that his theorem was not restricted to deterministic theories, and Clauser and Horne (1974) further developed this derivation of Bell's theorem. They showed that the available experimental data also falsified stochastic local theories, a conclusion, however, that depended on a supplementary "no-enhancement" assumption; which is weaker than the fair sampling assumption adopted by Clauser in the CHSH paper.[63] In any way, these results evidenced that it was locality and not determinism that was at stake in Bell's theorem.[64] Since then, to speak of the tests of

[61] Referee report on "Atomic Physics Experiments Using Lasers and Fast Atomic Beams", NSF proposal PHY-9016886, enclosed with Marcel Bardon [NSF] to F. M. Pipkin, 13 Dec 1990. Pipkin Papers [Accession 12802], box 21, folder "NSF awards."
[62] Clauser and Shimony (1978, pp. 1886-1900). The debate between Bell, on one hand, and Shimony, Horne, and Clauser, on the other hand, was published in *Epistemological Letters* and reprinted in Bell, Shimony, Horne, and Clauser (1985). Clauser and Shimony's paper became the canonical review on the experiments of Bell's theorem. It was cited 571 times, 64 until 1982. Wigner's 1970 paper included as a foonote a historical remark about what was von Neumann's main reason for stating the inadequacy of hidden variable theories. This footnote stirred up a strong Clauser's (1971a and 1971b) criticism, and the whole affair demanded the intervention of Shimony's diplomacy. Wigner (1971), and Wigner to Shimony, 5 Oct 1970, Wigner Papers, box 72, folder 1.
[63] Clauser and Horne's 1974 paper has 447 citations, 68 until 1982.
[64] The philosopher Karl Popper is an example of somebody who conjectured that the real conflict concerned determinism and not locality. For a criticism of this stance, see Bell (1972) and Clauser and Horner (1974, p. 526).

Bell's theorem as tests of determinism, as we have seen in the notice published in *Scientific Research,* is less than accurate.

## Settling the tie and turning the page

Let us now consider the period between 1975 and 1976. Indeed, it was a time of new and impressive experimental results as well of new experimental challenges concerning Bell's theorem. Two new experiments, by Fry and by Clauser, promise of a new one, by Aspect, new social settings for gathering the physicists involved in the debate, and the feeling, by certain physicists, of a turning point in this story were the main features of the time.

As we have seen, Edward Fry, from Texas A&M University, became interested in experiments on Bell's theorem in the early 1970s. Having learnt from his first unsuccessful application that many physicists had disdain for these experiments, Fry changed his strategy. Now, he tried funding from Research Corporation, and added two new assets. He attached two letters supporting the application, one from Eugene Wigner, who wrote "a strong supportive letter," and the other from his former adviser Bill Williams. And yet, he "tried to justify the pursuit of this subject in terms of a plethora of other more conventional capabilities that it also offered."[65] As a consequence of this strategy, Fry (1973) wrote that he was mainly interested in coincidence observation of optical photons from an atomic cascade as an experimental technique with variegated applications; and that experimental tests of local hidden variables were just one of the uses of such a technique, others being the determination of excited-state lifetimes and *g* values, branching ratios, absolute quantum efficiencies and source strengths.[66] The strategy adopted by Fry calls our attention to the kind of technical devices used in tests of Bell's theorem.

The experimental tests of Bell's theorem used technical devices that can be framed in what Peter Galison calls the "logic tradition" by contrast to those in the "image tradition." Both, according to Galison, formed two distinct traditions in the material culture of high-energy particle physics. In the logic tradition statistics plays the key role and there is no room for the "golden" and unique event that is important in the image tradition. The very counterintuitive nature of nonlocality prevents us from visualizing the phenomenon at stake. In the experiments with correlation of photons coming from atomic cascades, the most sensitive pieces of the apparatus are the photo detectors and the electronics to count them if there is coincidence, that is, if the pair of detected photons comes from the same cascade decay. Indeed, Galison's (1997, p. 464) logic tradition came "out of the established electronic logic tradition of counters," which was roughly available in the 1930s but was dramatically improved during the war.[67] As we have seen, Fry was particularly interested in improved detectors. Experiments with Bell's theorem also depended on good polarizers and optical filters in addition to efficient methods to excite the atomic samples to the right level in order to obtain the intended cascade decay. In contrast to Papaliolios, who used the polarizers produced by Polaroid Corporation, experimenters dealing with Bell's theorem used more traditional polarizers, like calcite prisms and "pile-of-plates" polarizers, a cluster of glass plates arranged in certain angles, due to their large efficiency in observing linear polarization. To excite the atomic sample, the experimenters used resonance absorption with radiation emitted by lamps and passed through interference filters or electron bombardment, methods that had the undesirable effect of exciting many levels and not only those which were intended. Among all these technical devices, the technical innovation that changed the scene of these experiments in the 1970s was a new technique to excite the atomic samples, the tunable dye laser. Peter Sorokin and John Lankard invented dye lasers in 1965, but "it took several years before tunability emerged as preeminent among its properties."[68] It quickly became a revolutionary technique for spectroscopy insofar as it yielded - within a certain range - the precise optical wavelengths one required to excite the atomic levels needed by experimenters.

---

[65] Edward Fry (personal communication, 5 Aug 2005).
[66] Fry's style was noted by Harvey (1980, p. 156) in the following terms: "… a major part of Fry's strategy was to develop experimental techniques per se, and then apply them to a number of quite different empirical problems."

Indeed, when Fry, helped by his graduate student Randall Thompson, went to carry out a new test of Bell's theorem he could take advantage of this new technique, the tunable dye laser. He used it for exciting exactly the atomic cascade of interest, and this permitted an improved speed of accumulating data. A number can summarize the improvements; while

---

[67] For the development of such techniques during the war, see Galison (1997, pp. 239-311).
[68] "The dye laser was also discovered independently by Mary L. Spaeth and D. P. Bortfield at Hughes and by Fritz P. Schaefer and coworkers in Germany. Both of these groups published later." Bromberg (1991, p. 184).

Clauser and Holt needed about 200 hours for collecting data, Fry and Thompson performed the experiment in 80 min. Their results strongly violated Bell's inequalities and matched quantum mechanical predictions. The Bell's inequality under consideration was $\delta \leq 0$, the quantum mechanical prediction was $\delta_{qm} = +0.044 \pm 0.007$, and they obtained $\delta_{exp} = +0.046 \pm 0.014$.[69]

Meanwhile, Clauser repeated at Berkeley, in slightly different conditions, the experiment carried out before by Holt and Pipkin at Harvard, with the aim of breaking the previous experimental tie. The main difference was due to economic and practical reasons. He used "pile-of-plates" polarizers instead of calcite prisms. He obtained results confirming quantum mechanical predictions and violating Bell's inequalities. The experiment ran for 412 hours. The Bell's inequality under consideration was $\delta \leq 0$, the quantum mechanical prediction was $\delta_{qm} = 0.0348$, and he obtained $\delta_{exp} = +0.0385 \pm 0.0093$.[70] For the second time, Clauser obtained experimental results that contradicted his hopes. He reported them to Wheeler in the following terms: "Dr. Henry Stapp here at LBL has told me that you were interested in the latest results from my experiments. These were attempting to reproduce the results observed by Holt and Pipkin at Harvard. Unfortunately, I have failed to do so, and obtained more or less good agreement with the quantum mechanical predictions."[71] Freedman and Holt also thought of repeating Holt and Pipkin's experiment, but this project did not happen.[72]

Clauser's and Holt's hopes and actual results deserve a remark on the literature in sociology of science. Harvey (1980, p. 157) convincingly argued that "the particular social, historical and cultural context in which the LHV [local hidden variable] experiments took place had a major effect on many features of these experiments." As examples of this effect, he cited location, physicists who carried them out, "the way in which they were presented, and the response to anomalous results." Holt's experiment fitted well in his claims. However, Harvey's claims were stronger, reflecting the then new trends in the sociology of science. For him (Harvey 1981, p. 106), "the social and cultural, as well as the technical, context in which a scientist finds himself will influence not only the style, timing and presentation of his work but also (at least in principle) its content." This strong claim about the content of the experimental results does not meet evidence when one compares the prospects nurtured by both Holt and Clauser. As Shimony (2002b, p. 74) remarked, after making the caveat that he "dislike[s] the idea that experimental results are theory laden, that somehow experimenters see what they want to see," the two physicists obtained results opposed to their expectations. Evidence about Clauser's hopes on violating quantum mechanics was already available at the time Harvey conducted his research. Harvey made reference to them, but did not extract the full consequences of the contrast between Clauser and Holt's cases. Nowadays, with more archival evidence available, it is harder to accept that local contexts determined the content of their experimental results.

---

[69] Fry and Thompson (1976). This paper received 154 citations, 56 until 1982.
[70] Clauser (1976). This paper received 126 citations, 57 until 1982.
[71] Clauser to John Wheeler, 27 Oct 1975. Clauser Papers.
[72] "You might be interested that I have decided to repeat Dick's mercury experiment here with pile-of-plates polarizers. [...] I hear you and Dick are considering collaborating on a similar repeat. Have you made a final decision on that?" Clauser to Freedman, 25 Jan 1974. Clauser Papers.

Since the early 1970s there already had been a small community of physicists who were interested in Bell's theorem. In the middle of the 1970s they looked to reinforce their links and to create opportunities for discussions and gathering. In addition to the usual trips and leaves of absence physicists use as regular means for circulating professional information, our protagonists used two others: The Erice Thinkshops on Physics, and the journal *Epistemological Letters*. The latter was an unusual vehicle for scientific debates. It was conceived as a permanent written symposium on "Hidden Variables and Quantum Uncertainty," and defined itself in this way: "*Epistemological Letters* are not a scientific journal in the ordinary sense. They want to create a basis for an open and informal discussion allowing confrontation and ripening of ideas before publishing in some adequate journal."[73] Indeed, the journal was more than this. It published short letters, kept open debates for several issues, announced news of interest, republished some papers, and even kept a list of the recipients of the journal. It was, in a certain sense, a predecessor of the contemporary Internet discussion lists. Instead of circulating via the electronic web it was mimeographed and sent to its recipients. Thirty-six issues were published from November 1973 to October 1984. About 60 authors wrote in the journal, and Shimony, Bell, d'Espagnat, Lochak, Costa de Beauregard, P. A. Moldauer, F. Bonsack, J. L. Destouches, and M. Mugur-Schaechter wrote at least five pieces each. Many papers were indeed published elsewhere but some debates which were not well documented elsewhere, such as the refusal of de Broglie and his collaborators to accept the full implications of Bell's theorem, are uniquely recorded there. It was published in Switzerland by the "Association F. Gonseth – Institut de la mèthode," under the editorial responsibility of the philosopher of science François Bonsack, who was the secretary of this association. Shimony also acted informally as an editor, publishing short reviews on the subject and actively intervening in the debates.[74] After the publication was over, he wrote a very favorable review of its existence:

> "The variety of the contributions and the vigor of the debates showed that the purpose was very well accomplished. Because of the brief time interval between issues and the absence of customary refereeing procedures, it was possible to carry on a debate more rapidly than in standard journals, and speculative ideas could be more easily made public. It is remarkable that in spite of the informality of *Epistemological Letters*, the typing of the articles, including mathematical formulae, was very accurate. The reputation of the written symposium spread rapidly, and many people throughout the world wrote to be added to the list of recipients." (Shimony, 1985)

---

[73] From the back cover of all *Epistemological Letters* issues. The University of Pittsburgh has a complete collection of this journal, a gift of Abner Shimony.

Erice, in Sicily, has been a favored destination for physics gatherings due to a conjunction of the natural and cultural appeal of the town and the restless initiatives of the Italian physicist Antonio Zichichi, head of its "Ettore Majorana Centre." Bell was close to Zichichi in their activities in high energy physics at CERN, and this relationship allowed him and d'Espagnat to organize the meeting, which took place in April 1976, and gathered together 36 participants from 25 laboratories and 9 countries.[75] This meeting had a double importance for the history of Bell's theorem. It was an environment for the socialization of some of the physicists involved in this research, and it was the stage for presenting and discussing Clauser's and Fry's yet unpublished results. It was instrumental for physicists who were established in the field, like Clauser, or entrant, like Aspect, to have a better socialization. Clauser (1992, p. 172) depicted the Erice meeting with these words, "the sociology of the conference was as interesting as was its physics. The quantum subculture finally had come 'out of the closet' and the participants included a wide range of eminent theorists and experimentalists." We shall see later its importance for Aspect. Socialization and professional recognition were important issues in the Bell's theorem saga, since recognition many times did not come in due time. We have much evidence concerning the lack of professional recognition even after the first experimental tests of Bell's inequalities.[76] It was not by chance that in this very meeting, Bell, highly sensitive to this issue as we have already seen, felt the need to criticize such an attitude and to present a rationale for pursuing the experiments. "The great success of quantum mechanics in accounting for natural phenomena in general will incline most people to expect it to remain successful here. Many people will even be intolerant of the idea of actually performing such experiments. […] We do a service to future generations by replacing *gedanken* with real experiments" (Bell, 1976, p. 440). If the first sentence concerns trust in quantum mechanics, the second is an open criticism of the still existing prejudices. The French physicist Franck Laloë (personal communication, 28 Mar 2005), a newcomer to the subject at that time, recalls, "being interested in the foundation of quantum mechanics was still considered sort of bad taste by most main stream physicists." However, none of this evidence is as telling as Clauser's case.

---

[74] Horne and Shimony (1973), Shimony (1980).
[75] A report of the conference, written by John Bell [Testing Quantum Mechanics], and the abstracts of the papers were published in *Progress in Scientific Culture – The Interdisciplinary Journal of the Ettore Majorana Centre*, 1/4, 439-460, 1976. I am grateful to Alain Aspect for sending me a copy of it.

Clauser's achievements in the 1970s were remarkable. He realized the full implication of Bell's theorem; carried out two key experiments on it, one of them being the first ever experimental result; and enhanced our understanding of the subject. In addition, he used the knowledge required for experiments with Bell's inequalities to contribute to the debate between supporters of semi classical radiation theories, notably Edwin Jaynes, and supporters of a full quantum treatment of radiation. He showed that experimental data from the polarization correlation of photons emitted in atomic cascades were incompatible with the semi classical theories and fitted well with quantum treatments (Clauser, 1972 and 1974). However, in spite of his achievements, Clauser faced hindrances achieving a professional career in physics based on experiments related to foundations of quantum mechanics. Indeed, he did not get a job.[77]

---

[76] My account strongly contrasts with Wick's stance (1995, p.244). He asked to some of these protagonists "if their participation in testing quantum mechanics adversely affected their standing among their peers, their ability to obtain research funding, or their job prospects, each replied simply 'no'."

[77] Although highly considered among both quantum opticians and physicists involved with foundations of quantum mechanics, Clauser eventually shifted his interests to other issues such as nuclear fusion, x-ray imaging, and recently Talbot-vonLau interferometry. One can conjecture about the role played in his decision by his feelings of lack of recognition of the subject among physicists, and the impact of such a lack on his own career.

The early 1970s were also the times of restrictions in the public funding of American science and this conjuncture brought consequences for the employment of the new physicists.[78] In addition to this background, to analyze why a researcher did not get a job in a number of institutions would require a more comprehensive study than the current paper. However, the available documentation shows that some physicists who decided the issue were influenced by the prejudice that experiments on hidden variables were not "real physics." Clauser faced both at Columbia and California the same hostile environment inherited from the 1950s. His former adviser, P. Thaddeus, wrote recommendation letters warning people not to hire Clauser if it is for doing quantum mechanics experiments, since it is "junk science" (Clauser 2002b, p. 12). In fact, some of his possible employers thought the same, in spite of the many letters of recommendation he had.[79] Shimony reported to him that "when I saw d'Espagnat last week he had a letter from the Dep't Chairman at San Jose, inquiring whether what you have been doing is real physics. Needless to say, he'll write a strong letter answering the question in your favor. I'm sorry, from that evidence, to find that your job situation is still unsettled."[80] Retrospectively, Clauser (2002b, p. 18) admits that he was not smart enough not to give talks on Bell's inequalities while looking for a job. "I was sort of young, naïve, and oblivious to all of this. I thought it was interesting physics. I had yet to recognize just how much of stigma there was, and I just chose to ignore it. I was just having fun, and I thought it was interesting physics. I was just trying to understand what was going on." Shimony, also retrospectively, suggests a sociological explanation for this fact; "certainly there was a lot of interest [in Bell's theorem], but that doesn't mean there was enough to get majority votes in physics departments to bring somebody whose main credentials were an experiment concerning hidden variables."[81] Independent of his admitted naiveté and the job crisis in American physics, it is interesting to record how some leading physicists reacted to the physics he was doing. Clauser (2002a, p.71) reports that while he "was actually performing the first experimental test of the CHSH-Bell predictions as a postdoc at UC-Berkeley, [... he] made an appointment with Prof. Richard Feynman to discuss these same questions. Feynman was very impatient with [him]." Feynman's stance was: "Well, when you have found an error in quantum-theory's experimental predictions, come back then, and we can discuss your problem with it."[82]

---

[78] According to Kevles (1978, pp. 421-423), this period represented "a degree of disestablishment" in American physics. It had begun in the middle of the 1960s, signalling the end of the "post-Hiroshima honeymoon," but by the early 1970s "the cutbacks […] created an employment squeeze reminiscent of the 1930s depression."

[79] "What is the situation regarding employment next year? If any more letters should be written, let me know;" Abner Shimony to John Clauser, 19 May 1972. "In reply to your letter of January 9, I am happy to write in support of Dr. John Clauser as a candidate for a faculty position at UCSC. I believe he shows promise of becoming one of the most important experimentalists of the next decade. […] I say these things in spite of the fact that Clauser's results spell trouble for my own pet theory." Edwin T. Jaynes to Peter L. Scott [Chairman, Board of Studies in Physics, University of California at Santa Cruz], 31 Jan 1973. Clauser Papers.

[80] Shimony to Clauser, 8 Aug 1972, Clauser Papers.

   The second important feature of the Erice meeting is related to the fact that it was considered a turning point towards the recognition of quantum nonlocality as a physical effect. An evidence of the high expectative the meeting awakened is the fact that Bell had especially invited the reputed University of Chicago's experimentalist, Valentin Telegdi, to give his opinion on the recent experiments performed by Clauser and Fry. Telegdi had not had any previous involvement with experiments in foundations of quantum mechanics. Bell and d'Espagnat also invited Pipkin, responsible for the only diverging result, in order to bring together the main protagonists of this story. "It will be good if our meeting can contribute to getting to the bottom of the differences between he various experiments. In any case it will be a great pleasure to see you in Erice," wrote Bell to Pipkin.[83] In his concluding remarks, Bell (1976, p. 442) was sober, "such atomic cascade experiments were reported by Clauser, Pipkin, and Fry (a very elegant new experiment). Three of the four experiments are in excellent agreement with quantum mechanics. But that of Holt and Pipkin is in serious disagreement. [...] After discussions at this meeting it remains unknown what, if anything, went wrong in the Holt-Pipkin experiment." Bell's sobriety was not widely shared. The French physicist Olivier Costa de Beauregard reported this feeling in *Epistemological Letters*: "Les tout derniers résultats expérimentaux, non encore publiés [...] sont

---

[81] This conjecture, however, does not attenuate Shimony's criticisms: "I think he was treated very shabbily. He's a brilliant man, a very good experimenter, and really a good theoretician also." Shimony (2002b, 9 Sep 2002, p. 82-3).
[82] Feynman's opinion on Bell's theorem deserves further research to track its evolution. His views after Aspect's experiments, in 1984, will be commented later. After the talk reported by Clauser, while visiting Texas A&M at Austin, in 1974, Feynman was approached by Edward Fry and James McGuire to discuss their planned experiment, and reacted positively. Edward Fry (personal communication 5 Aug 2005).
[83] John Bell to Francis Pipkin, 22 Dec 1975, Pipkin Papers, accession 13034, box 2, folder "Correspondence Jan – Apr – 1976."

*explicitement en faveur* de la Mécanique Quantique, et *confirment donc la réalité du paradoxe*. [84] Indeed, the audience sensed that one page in the hidden variable story was being turned. The impression was that from now on one could speak of a new quantum physical effect, quantum non-locality, as Lalöe testifies:

> "This meeting coincided with an important turn in physics. Until about that time it had been possible to believe that the Bell inequalities were obeyed by Nature, since they relied on very general assumptions (very much in the spirit of relativity). […] For some time, in particular in view of the experiments performed by Pipkin at Harvard University, some doubt remained indeed possible. But when John Clauser and his group gave their results, and then even more when Ed Fry came with another series of even more precise experiments, the agreement between the results and quantum mechanics was so impressive that no-one could anymore still think seriously that the Bell inequalities were obeyed by physics." (Lalöe, personal communication, 28 Mar 2005)

Bell, Costa de Beauregard, and Lalöe's words acquire more vivid colors when compared with the following review of experiments on hidden variables written two years before by three experimentalists who were involved with the subject, Freedman, Papaliolios, and Holt.[85] After revising the results obtained by Freedman and Clauser, by Holt, and by experiments on annihilation of positrons, they concluded:

> "We note, in conclusion, that the problem of the validity of local hidden variable theories rests with the experimentalists. New experiments are in progress or are being planned by several groups and we can hope for a solution in the near future. It is fair to say that the existing evidence still favors quantum mechanics; nevertheless, the question is of fundamental importance and there is too much at stake to allow any experimental discrepancy to remain unexplained." (Freedman, Holt, and Papaliolios, 1976, 57).

---

[84] O. Costa de Beauregard – Nouvelles du colloque sur le Paradoxe EPR au Centre Ettore Majorana, à Erice, 18-23 avril 76, *Epistemological Letters*, 10[th] issue, p. 26, May 1976.
[85] The paper was written in 1974, by initiative of Freedman and Holt, for a volume in honor of Louis de Broglie. Freedman to Papaliolios, 18 Mar 1974, Holt to Papaliolios, 1 Apr 1974, Papaliolios Papers, box 23, folder "Hidden Variables – Paper in honor of de Broglie."

However, despite of Fry and Clauser's results favoring quantum mechanics, Bell did not consider the history over, and in his final report in fact he set a new agenda with a new experimental challenge. He resumed an idea due to David Bohm (1951, p. 622), "that [in a EPR experiment] while the atoms are still in flight, one can rotate the apparatus into an arbitrary directions." As in the real experiments the analyzers are set before the experiment one cannot discard that the existence of an unknown subluminal interaction between the analyzers is responsible for the correlations. If experiments confirm the existence of such an interaction, Bell's condition of locality, on which Bell's theorem depends, would loose its validity. According to Bell (1976, p. 442), "now it can be maintained that the experiments so far described have nothing to do with Einstein locality It is therefore of the very highest interest that an atomic cascade is now under way, presented here by Aspect, in which the polarization analyzers are in effect *re-set while the photons are in flight.*"

**New challenges: "while the photons are in flight"**

Alain Aspect's road to experiments in foundations of quantum mechanics reflects a transitional time rather than Shimony's and Clauser's roads. It was a transitional time from the consideration of a subject as a philosophical quarrel to its recognition as an interesting field of physical research. His biographical profile not only reflects that time but also evidences his active contribution to awarding these issues the respectability they deserved. Born in 1947, Aspect did his undergraduate studies at École normale supérieure de l'enseignement technique (ENSET), in Cachan, south of Paris, while taking physics courses at Orsay. Next, he did the French *doctorat de troisième cycle*, on holography, also at Orsay. Soon after his work on holography, he became disappointed with physical research, and planned to become a teacher, having successfully ranked in 2[nd] in the French national contest named "aggregation" (Aspect, personal communication, 28 Feb 2005). Following the French tradition of replacing military duties with civil service, he went to Cameroon, teaching there between 1971 and 1974. However, while in Africa, his plans changed once more. He realized that just teaching would become boring in the medium term, and he studied the new textbook on quantum mechanics written by Claude Cohen-Tannoudji, Bernard Diu, and Franck Laloë (1973), which revived his interest in physical research, especially in optics.[86] Coming back to

---

[86] "I can say that my previous studies in quantum physics had been totally disappointing: it was just solving partial differential equations about 'rigid rotator' and so, not physics

France, Aspect got a tenured position at ENSET, his former school; a job which gave him freedom to choose his themes of research. In the fall of 1974, interested in resuming research but on subjects in which the "quantum weirdness" appears, he headed to the Institut d'Optique at Orsay, to talk with Christian Imbert, a young professor who had done experiments related to the photon self-interference, looking for his French *doctorat d'état*. Imbert, who was in touch with Bernard d'Espagnat and Olivier Costa de Beauregard, handed him a bibliography related to Bell's inequalities, "to look into that for a possible subject."

Aspect (ibid.) became fascinated reading Bell's paper, realized the conflict between the experimental results of Clauser and Freedman, on one hand, and Holt and Pipkin on the other hand, and found the right subject for his dissertation: "an experiment in which the polarizers would be rotated while the photons were in flight." Aspect's project would resonate with the interest of some of our protagonists. André Maréchal, then the director of the Institut d'Optique, asked d'Espagnat for a good subject for the dissertation of a good researcher who had come back from Africa, i.e. Alain Aspect. d'Espagnat (2001, p. 11) discussed the subject with Bell and they agreed to propose to Aspect that he settle the conflicting results obtained in the previous experiments, which at the time were in a tie between Clauser's first experiment and the one by Holt and Pipkin. Aspect also realized that this experiment should be a long term project due to its technical challenges and the need of getting acquainted with the subject, because he had had no previous training in modern experimental atomic and laser physics. However, if the intrinsically long *doctorat d'état* and his tenured position at ENSET could be accommodated by this project, he needed advisers and funding. At this point, a series of conversations involving Bell, d'Espagnat, Imbert, Costa de Beauregard, in addition to letters from Arthur Wightman and Alfred Kastler supporting the funding of such project, sealed favorably the fate of the project.[87] However, it should be noted that the French leaders in atomic and laser physics were not initially involved with Aspect's experiment, since that leadership was shared by Jean Brossel, who was the head of according to my view of physics. The textbook of CCT et al totally changed my view on that," Aspect (ibid.).

[87] Aspect (personal communication, 28 Feb 2005). The acknowledgments in Aspect's (1976) proposal of experiments evidence the patronage for them: "the author gratefully acknowledges Professor C. Imbert and Dr. O. Costa de Beauregard for having suggested this study and for many fruitful discussions. He especially thanks Dr. J. S. Bell for his encouragement, and Professor B. d'Espagnat for his thorough consideration and discussion of the theoretical aspects of our scheme."

the laboratory at Ecole Normale Supérieure in which Kastler, Cohen-Tannoudji, and Lalöe were working, and Pierre Jacquinot, head of the Laboratory Aimé Cotton.

At the beginning, Aspect worked alone. He learned about the measurement of "photon coincidence at the electronic shop of the CEA at Saclay," borrowed equipment from them, and talked to experimentalists at Ecole Normale Supérieure and Laboratory Aimé Cotton, who were using tunable lasers and atomic beams. He also learned how to speak on Bell's inequalities to people who were not a priori interested in foundations of quantum mechanics. According to him, "I discovered that if I presented things in a very simple and naïve way, just as I had understood Bell's paper the first time I had read it, most of the public a priori skeptic (not to say more) would become interested and sympathetic."[88] Nowadays, people who attend Aspect's lectures consider he became a charismatic lecturer. Things evolved positively. In the 1976 Erice meeting, in addition to meeting physicists already involved with Bell's theorem, he met Laloë,[89] who convinced Cohen-Tannoudji that Bell's inequalities were an interesting physics topic, and introduced Aspect to him. According to Aspect, the interaction with Cohen-Tannoudji produced a "phase transition" in the way he was regarded by his colleagues around 1978-1979, when collaboration with Cohen-Tannoudji's team on a side subject, different from Bell's inequalities, allowed him to establish close intellectual links with this highly respected French physicist.

In the late 1970s, signs appeared suggesting that Bell's theorem was gaining wider recognition. In July 1979, Bell (1980) was one of the invited speakers at the Conference of the European Group for Atomic Spectroscopy, to talk on "Atomic-cascade photons and quantum-mechanical nonlocality." In June 1980, the Collège de France, under the patronage of Cohen-Tannoudji, who was then at the apex of French physics, with a chair in this prestigious French institution, organized an international colloquium on the conceptual implications of quantum physics in which Bell's inequalities and Aspect's ongoing experiments were presented.[90] The intense experimental and theoretical scientific activities on Bell's theorem awakened interest beyond physics. d'Espagnat (1979) published in *Scientific American*, the renowned American popular science magazine, a non-technical account of Bell's theorem and its experimental tests. In the late 1970s, the new trend in the sociology of science, interested as it was in the study of scientific controversies, produced a number of papers dedicated to the debate on hidden variables in quantum mechanics.[91]

---

[88] Aspect (ibid.). As examples of his approach, see Aspect (1983 and 2002).

[89] "Alain is right when he mentions our many friendly discussions at that time and later, during his thesis in Orsay. […] Sometimes he was explaining to me things that I had not understood, sometimes it went the other way. For instance I remember that one day in Paris I suggested to him the two channel experiment (with birefringent filters), which turned out to be a good idea, but certainly not for the reasons I was proposing, which were incorrect!" Lalöe (ibid.).

Aspects' experimental results eventually came out, between 1981 and 1982. The initial project of a single experiment had become three. In collaboration with his undergraduate student Philippe Grangier and the research engineer Gérard Roger, he once again performed the test of Bell's inequalities, as Clauser and Fry had done, but used two tunable lasers to excite the sample, providing him a source with higher efficiency. The counting lasted 100 seconds. In addition, as the Paris group separated source and polarizer by 6.5 m, they could rule out Furry's conjecture, which suggested that the quantum nonlocality would vanish after the photons travel a distance of the order of the coherence length of their associated wave packets; which meant a distance of 1.5 m in this experiment. In mathematical terms, a pure state would evolve towards a mixture of factorizing states. With the same collaborators Aspect used two-channel polarizers, which allowed a straightforward transposition of EPR *gedanken* experiment. In previous experiments when one of the detectors is not triggered, one could not know whether it was a result of the low efficient detectors or whether the polarizer has blocked the photon, which would be a real measurement. For this reason auxiliary experiments with the polarizers removed were needed to circumvent the intrinsic deficiency of the setup. Finally, with Jean Dalibard and Gérard Roger, he produced the first test of Bell's inequalities with time-varying analyzers. Aspect's ingenuity was to use a switch to redirect the incident photons to two different polarizers. This device works through an acousto-optical interaction with an ultrasonic standing wave in water.[92] All the experimental results violated Bell's inequalities and strongly confirmed

---

[90] The lecturers at the Collège de France were Laloë, Bell, Aspect, Shimony, and d'Espagnat, and the conference brought together 25 people, both physicists, and philosophers; *Journal de physique*, Tome 42, Colloque C2, 1981.
[91] Pinch (1977), Brush (1980), and Harvey (1980 and 1981).
[92] Aspect, Grangier, and Roger (1981 and 1982); Aspect, Dalibard, and Roger (1982). The first paper received 409 citations, the second 541, and the third 835 citations. The choice for this switch, instead of using Kerr or Pockels cells as first thought by Clauser, was determined by the consideration that with these cells "only very narrow beams could be transmitted, yielding very low coincidence rates; as these cells heat up, and then become inoperative, long

quantum mechanics' predictions. In the first experiment the Bell inequality was $\delta \leq 0$, the quantum mechanical prediction was $\delta_{QM} = 5.8 \times 10^{-2} \pm 0.2 \times 10^{-2}$, and the experimental result was $\delta_{exp} = 5.72 \times 10^{-2} \pm 0.43 \times 10^{-2}$, which violated the Bell inequality by more than 13 standard deviations. For the second experiment, the Bell inequality at stake was $-2 \leq S \leq 2$, $S_{QM} = 2.70 \pm 0.05$, and the result was $S_{exp} = 2.697 \pm 0.015$, to that date the strongest violation of Bell's inequalities ever reported. The third experiment is telling by what it was measuring, a Bell's inequality using time-varying analyzers, and due to this reason it was the result that most resonated in the physics community, but its accuracy was less than the previous ones. It tested Bell's inequality $S \leq 0$, $S_{QM} = 0.112$, and the experimental result was $S_{exp} = 0.101 \pm 0.020$, violating Bell's inequality by 5 standard deviations in runs which lasted 200 minutes.

Aspect's experiments made his professional reputation, and his ongoing research on new fundamental phenomena, like experiments with just one photon, with laser cooling below the one photon recoil, and, more recently, on the Bose - Einstein condensate, carried him to a position of leadership in quantum optics on the world stage, while in France his prestige can be evaluated by his 2001 election to the French *Académie des sciences*.

In 1982 Alain Aspect was one of the invited speakers at the Eighth International Conference on Atomic Physics, held in Sweden, to report on his experiments on Bell's inequalities. The American physicist Arthur Schawlow, Physics Nobel Prize winner in 1981, was requested to make the final report of the conference. He chose Bell's theorem and its experiments as the main topic of his speech:

"Physical metaphors, such as the dual concepts of particles and waves in dealing with the light and atoms, are more than just conveniences, but rather are practical necessities. [...] But the experiments on Bell's inequalities are making it difficult for us to continue using some of our familiar physical metaphors in the old ways. We are used to thinking that light waves are produced at an atom with definite polarizations and are subsequently detected by remote detectors. However, the experiments show that if anything is propagated, it seems to convey more polarization information than a transverse wave. [...] As an experimentalist, I like to think that there is something

---

runs would be prohibited." In addition, the calibration of the system would be "exceedingly difficult" due to the need of monitoring the change of the polarizer orientations. Aspect (1976, p. 1945).

there that we call an atom, and that we can make good measurements on it if we careful not to disturb it too much. But the experiments on polarization of correlated photons don't bear out these expectations." (Schawlow, 1983)

Two years later, Feynman, who once refused to talk about hidden variables with Clauser while the first experiment was being carried out, attended a seminar given by Aspect at Caltech on the tests of Bell's theorem and wrote to him, "once again let me say, your talk was excellent." At this seminar, Aspect finished his talk by quoting a certain paper whose author derived results similar to Bell's inequalities and went on to discuss whether it was "a real problem." According to Aspect, this author gave an answer so unclear that he "had found it amusing to quote it as a kind of joke to conclude this presentation." Only at this point, did Aspect reveal the name of the author, Richard Feynman. According to Aspect, nobody in the audience laughed until Feynman laughed. Feynman checked the quotation and wrote to Aspect conceding he was right.[93] We conclude this section with this emblematic anecdote; the story of the recognition of Bell's theorem is over.

**Conclusion**

Hidden variables, considered a philosophical matter thirty years before, entered the optics laboratories, and occupied a place in mainstream physics. This is not to say that physicists reached a full consensus on the meaning of these experiments and that philosophical issues vanished.[94] Indeed, the compatibility between quantum physics and its non-locality, on one hand, and special relativity, on the other hand, remained a matter of dispute. There appeared demonstrations that quantum mechanics cannot be used to exchange superluminal messages. Bell, himself never felt comfortable with this kind of compatibility between these two major physical theories.[95] Other physicists did trust that detectors with higher efficiencies would lead to violations of quantum mechanics. Indeed, one year after the publication of Aspect's experiments, T. W. Marshall, Emilio Santos, and Franco Selleri (1983), published a paper entitled "Local realism has not been refuted by atomic cascade experiments," in which they built a hidden-variable model able to mimic quantum-mechanical predictions. Marshal, Santos, and Selleri had resumed a line of approach first suggested by Clauser and Horne (1974), by relaxing the fair sampling assumption adopted in the CHSH 1969 paper. The analysis of this stance and its resonance among physicists, however, falls beyond the time table of this paper.

---

[93] Richard Feynman to Aspect, 28 Sep 1984. R. P. Feynman Papers, box 22, folder 15. Aspect (personal communication, 20 Apr 2005). Feynman's quotation, in Feynman (1982, p. 471), is: "It has not yet become obvious to me that there's no real problem. I cannot define the real problem, therefore I suspect there's no real problem, but I'm not sure there's no real problem. So that's why I like to investigate things." Before this fragment, Feynman had written, "Might I say immediately, so that you know where I really intend to go, that we always have had a great deal of difficulty in understanding the world view that quantum mechanics represents."

[94] An example of a good physicist who did not grasp the full meaning of Bell's theorem even after Aspect's experiments is Abraham Pais. While writing Einstein's biography, Pais (1982, ch. 25c) assessed the EPR paper had no bearing on physics and did not cite Bell's theorem as a development of this issue.

The most important historical lesson from the period analyzed, I think, is that the path from philosophy to physics required not only good theoretical ideas, experimental skills, and technological improvements, but also a change in the physics community's attitude about the status of the foundations of quantum mechanics as a subject for physics research. However, the path from philosophy to physics was slow and sinuous, involved diverse factors, and not only the ones I discussed here. Even on the road I presented here, the perceptions of its protagonists about the recognition of these achievements evolved in different ways, according to their personal experiences and local contexts. The 1970s were a transitional decade for the research on Bell's theorem, in particular, and for the foundations of quantum mechanics, in general. The role of local contexts and personal stories can be measured if one considers that while Shimony met at the 1970 Varenna meeting a small but supportive environment for his research and Holt did not feel prejudices against the subject of his dissertation at Harvard in the early 1970s, Clauser only found a similar environment at the 1976 Erice meeting and Aspect only felt himself well accepted in the main milieu of the French Optics by 1978.

In parallel with the differences we have seen in this paper, some common features can be extracted from the biographical sketches of our characters. A rough collective biography of them can be drawn noting that many of them were after all, dissenters, or quantum dissenters. They fought against the dominant attitude among the physicists according to which foundational issues in quantum mechanics were already solved by the founding fathers of the discipline. Some of them, such as Bell, Clauser, and Shimony, were hard critics of the complementarity interpretation. The common ground of the quantum dissenters was minimal and focused just on the importance of the research into the foundations of quantum mechanics. They supported different interpretations of this physical theory and chose different approaches and issues in their research. While Clauser did not trust in the Copenhagen interpretation, Aspect had no philosophical quarrels with this interpretation in itself; while in the 1950s Bohm tried to build models for mimicking quantum mechanics, Bell gave his attention to the critical analysis of its assumptions. The fact that their common platform was the critical analysis, both theoretical and experimental, of the foundations of quantum physics, instead of the development of just one alternative interpretation, or even the advocacy of their philosophical credo, was one of the sources of their strength. They had the benefit of a new professional environment, since the old generation of the founding fathers of quantum physics were no longer in the field of combat. They were also benefited by cultural changes in the late 1960s, which opened room for criticisms of science and criticisms within science. And yet, they did not just reflect their times but they also contributed to changing them. They were led to these issues not only by scientific motivations; philosophical, pedagogical, and even political factors were also influential, while varying from case to case. Their story is as a whole a story of success since foundations of quantum mechanics, or at least some research from this field, entered the mainstream of physics. Nevertheless, in each individual case, recognition did not always come in due time, and that may explain their different appreciation of the evolution of the field to which they dedicated their energy.

---

[95] Eberhard (1978), Ghirardi, Rimini, and Weber (1980), Aspect (1981), Page (1982), and Bell (2004).

## Acknowledgement

An early version of this paper was presented as an invited talk at the session "Quantum Optics Through the Lens of History," Forum of History of Physics, American Physical Society, April 16 – 19, 2005, Tampa, FL. I am grateful to Joan Bromberg for the invitation, comments, and her help in editing this text. I wrote this paper as a Senior Fellow at the Dibner Institute for the History of Science and Technology, at MIT, and Visiting scholar at Harvard University. Grants from CNPq & CAPES (Brazil) have partially funded this research. I am also indebted to Abner Shimony, Olivier Darrigol, David Kaiser, Stefano Osnaghi, Martha Bustamante, Edward Fry, Alain Aspect, and an anonymous referee of this journal for their comments on previous versions of this paper. I am also grateful to the following institutions for authorization to quote from their archives: Harvard University Archives [Costas Papaliolios Papers, Francis M. Pipkin Papers – Accessions 12802 and 13034]; California Institute of Technology Archives [Richard P. Feynman Papers]; Niels

Bohr Archive [Léon Rosenfeld Papers]; Manuscripts Division - Department of Rare Books and Special Collections, Princeton University Library [Eugene Wigner Papers]; Institue Archives, MIT [Norbert Wiener Papers – Collection MC022]; and American Institute of Physics [interviews with Abner Shimony, Bernard D'Espagnat, and John Clauser]. I consulted the John Francis Clauser personal papers due to kindness of Joan Bromberg, who provided me copies of them.